\documentclass[journal]{IEEEtran}
\usepackage{csquotes}
\usepackage{ragged2e}
\usepackage{graphicx}
\usepackage{epstopdf}
\usepackage{float}
\usepackage{subcaption}
\usepackage{amsthm}
\usepackage{color}
\usepackage{tabulary}
\usepackage{verbatim}
\usepackage{enumerate}
\usepackage{rotating}
\usepackage[mathscr]{eucal}
\usepackage{ifthen}
\usepackage{makeidx}
\usepackage{multirow}
\usepackage{amsmath} 
\usepackage{amssymb}
\usepackage{booktabs}
\usepackage{textcmds}
\usepackage{nomencl}
\usepackage{verbatim}
\usepackage{rotating} 
\usepackage[mathscr]{eucal}
\usepackage{ifthen}
\usepackage{makeidx}
\usepackage{multirow}
\usepackage{amsmath} 
\usepackage{enumerate}
\usepackage{amsfonts}
\usepackage{amsmath}
\usepackage{hyperref}
\usepackage{algorithmic}
\usepackage{color,soul}
\usepackage[left=0.67 in,top=0.67 in,right=0.67 in,bottom=0.67 in]{geometry}
\usepackage[normalem]{ulem}
\usepackage{cite}
\usepackage{cases}
\usepackage{nomencl}
 
\usepackage[font=footnotesize]{caption}
\usepackage[font=footnotesize]{subcaption}
\renewcommand\IEEEkeywordsname{Index Terms}

\renewcommand{\figurename}{Fig.}

\usepackage{algorithm}
\usepackage[table]{xcolor}
\usepackage{algorithmic}
\usepackage{color}

\newcounter{MYtempeqncnt}
\usepackage{multirow}
\usepackage{hyperref}
\hypersetup{
    colorlinks=true,
    linkcolor=blue,
    filecolor=magenta,      
    urlcolor=cyan,
}
 
\usepackage{pdfpages}
\pdfoutput=1
\usepackage{fancybox,framed}
\ifCLASSINFOpdf
\else
\fi
\begin{document}
\begin{titlepage}
\centering
\doublebox{%
\begin{minipage}{6in}
\begin{center}
This is the accepted version
\end{center}

\textbf{Live link to published version in IEEE Xplore:} \url{https://ieeexplore.ieee.org/document/8632732}
\newline
\newline
\textbf{Citation to the original IEEE publication:} A. Thakallapelli, and S. Kamalasadan, \enquote{Alternating Direction Method of Multipliers (ADMMs) Based Distributed Approach For Wide-Area Control}, IEEE Transactions on Industrial Applications, Volume: 55, No: 3, pp. 3215 {-} 3227, Feb. 2019.
\newline
\newline
\textbf{Digital Object Identifier (DOI): 10.1109/TIA.2019.2896837}
\newline
\newline
The following copyright notice is displayed here as per Operations Manual Section 8.1.9 on Electronic Information Dissemination (known familiarly as "author posting policy"):
\newline
\newline
\textcopyright{ 2019 IEEE}. Personal use of this material is permitted. Permission from IEEE must be obtained for all other uses, in any current or future media, including reprinting/republishing this material for advertising or promotional purposes, creating new collective works, for resale or redistribution to servers or lists, or reuse of any copyrighted component of this work in other works.
\end{minipage}}

\end{titlepage}
%
\title{Alternating Direction Method of Multipliers (ADMM) Based Distributed Approach For Wide-Area Control}
%
%
%

\author{Abilash~Thakallapelli,~\IEEEmembership{Student~Member,~IEEE,}
        and~Sukumar~Kamalasadan,~\IEEEmembership{Senior~Member,~IEEE}
\thanks{A. Thakallapelli and S. Kamalasadan (corresponding author), are with the Power, Energy and Intelligent Systems Laboratory, Energy Production Infrastructure Center (EPIC) and Department of Electrical Engineering, University of North Carolina at Charlotte, Charlotte, NC 28223 USA (e-mail: athakall@uncc.edu, skamalas@uncc.edu).}
}

\maketitle

\begin{abstract}
 In this paper, an ADMM based novel distributed wide-area control architecture is proposed for damping the inter-area oscillations. In this approach, first, an interconnected power system is divided into areas based on coherency grouping. Second, local processors are assigned on each area that estimates a black-box transfer function model based on Lagrange multipliers using measurements. These local area processors are then used to estimate a global transfer function model of the power system based on a consensus algorithm through a global processor. After convergence, a transfer function residue corresponding to the inter-area mode of interest is derived, to determine optimal wide area control loop. Finally, a wide-area damping controller is designed based on this information. The efficacy of the controller is validated using two area and IEEE-39 bus test systems on RTDS/RSCAD® and MATLAB® co-simulation platform.
\end{abstract}

\begin{IEEEkeywords}
Damping Controller, Interarea Oscillations, Wide-area Control, Alternating Direction Method of Multipliers (ADMM).
\end{IEEEkeywords}

%
\IEEEpeerreviewmaketitle

\section{Introduction}
%
%
%
%
\IEEEPARstart{F}{or} reliable operation, electro-mechanical oscillations which arise in large interconnected power systems due to disturbance should be damped in a timely manner. The electro-mechanical oscillations of generators with respect to remaining part of the system are called local modes, whereas groups of generators oscillating together against other groups through the tie-lines are called inter-area modes \cite{refa1}. The frequency of inter-area oscillations is between 0.1-1.0 Hz. Higher penetration of renewable energy resources and unpredictable loads makes the inter-connected power systems to operate close to limits. This condition increases the stress on the power system and can deteriorate the inherent damping of the system. Thus inter-area oscillation damping is even more critical and difficult in the modern power grid.  Unfortunately, the local measurements damping influence on the inter-area modes are comparatively less when compared to non-local measurements \cite{refa2,07206613}, so the effectiveness of conventional power system stabilizer (PSS) in damping of inter-area modes is thus limited. 

Several methodologies to identify and damp inter-area oscillations considering optimal wide area control loop (input/output signal selection) are reported in the literature. This include but not limited to, residue analysis \cite{refresidue1,refresidue2,refresidue3}, relative gain array (RGA) \cite{refrga1}, combined residue and RGA method \cite{refcom1}, and geometric measure of joint observability/controllability \cite{refgeo1,refa4}. However, these methods are formulated using state-space matrices obtained through linearizing the system at a particular operating point and analyzing modes at that point. This approach is not feasible as power system is non-linear and dynamic in nature. Thus, the control loop identified by linearizing at an operating point may not be effective in damping inter-area modes. More recently, in addition to state-space methodologies, converter based devices are used for wide-area control. For example, in \cite{07378520}, the measure of observability/controllability was used to design a coordinated WADC considering WTGs in the control process. A robust WADC design using HVDC links  is proposed in \cite{07811175} and sliding mode based damping control for DFIG is proposed in \cite{07529091}. However, these methods did not discuss the optimal signal selection and feasibility of online implementation.

To overcome drawbacks of linearization based signal selection, measurement-based methods have been designed. In these methods, measurement data is analyzed with a centralized controller to identify the optimal control loop. Since measurements are updated as the system operating condition changes, these methods keep track of changing operating conditions while identifying optimal control loop. Several centralized methods using measurements were reported in literature such as 
Hilbert–Huang transform \cite{refa6}, subspace identification technique \cite{refsub1}, energy function based \cite{refa7}, mode metering \cite{refa9}, and principal component analysis (PCA) \cite{refa10}.
A combination of state-space and measurement based technique for a wide-area control strategy to synthesis control set point for HVDC/FACTS is presented in \cite{07206613}. The  application of Kalman Filtering technique to estimate the inter-area modes is discussed in \cite{07268768}. Using this information a  controller is designed for SVC(Static VAR Compensator).  In \cite{07394940} a parametrically robust wide area damping controller considering WTG is proposed. Most of these methods are centralized and did not address the problem of optimal wide-area control loop identification. For large scale network the centralized approach for wide area signal selection may not be feasible due to various factors like data volume, data transfer capability, computational time etc. 

Considering these factors and to overcome the drawbacks of earlier methods reported in the literature, this paper introduces a novel method to identify optimal wide area control loop for wide area damping controller (WADC) using distributed algorithms. The objective of the distributed algorithm is to break a problem into sub-problems such that each sub-problem can be solved in parallel. One such simple algorithm which can solve sub-problems in parallel by sharing relatively small packet of messages is ADMM \cite{refadmm}. Thus, a methodology based on ADMM for WADC is proposed. 

This paper is an extended version of \cite{refconf}. For the proposed study, initially the interconnected power system is divided into areas based on coherency grouping of generators. Then each local area processor is designed to  estimate multi-input-multi-output (MIMO) black-box transfer function model based on Lagrange multipliers method using measurements. The local area processors are then tasked to communicate with the global processor to estimate a global transfer function model of the power system. Upon convergence, the residue corresponding to inter-area mode of interest obtained from the estimated global transfer function is used to identify optimal wide area control loop. Further, information of residue and corresponding eigenvalue is used for WADC design. The effectiveness of the proposed optimal control loop selection methodology and wide area controller design is validated using two area and IEEE-39 bus test systems on RTDS/RSCAD® and MATLAB® co-simulation platform.

The major contributions of this work are:
\begin{description}
   \item[$\bullet$] An algorithm to solve power system MIMO transfer functions using a novel approach based on ADMM is designed. This algorithm changes with operating conditions and provides a dynamic system model.
  \item[$\bullet$] A new method for online selection of optimal wide-area control loop using ADMM techniques is developed. The approach is adaptive and can damp multiple oscillations in the power grid.
  \item[$\bullet$] A novel experimental test-bed for power system wide-area monitoring techniques using distributed algorithms is developed.
  \item[$\bullet$] The proposed approach provides significant damping improvement (overall improvement of 60-80\% in damping) when compared to the existing methods. 
\end{description}

The rest of the paper is organized as follows: Section II discusses the proposed approach and problem formulation. In section III, the experimental setup for implementing the proposed signal selection method is discussed. In section IV, implementation test results are illustrated and Section V concludes the paper.

\section{Proposed Approach and Problem Formulation}  
\begin{figure}[!h]
\centering
\includegraphics[width=0.475\textwidth]{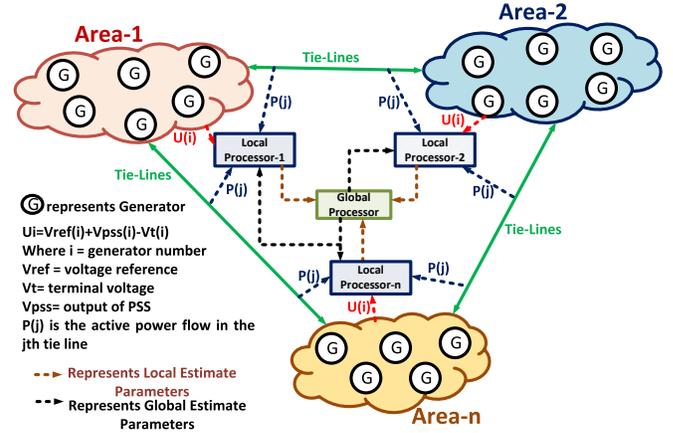}
\caption{Proposed distributed architecture for wide area control.}
\label{fig1}
\end{figure}

\begin{figure}[!h]
\centering
\includegraphics[width=0.375\textwidth]{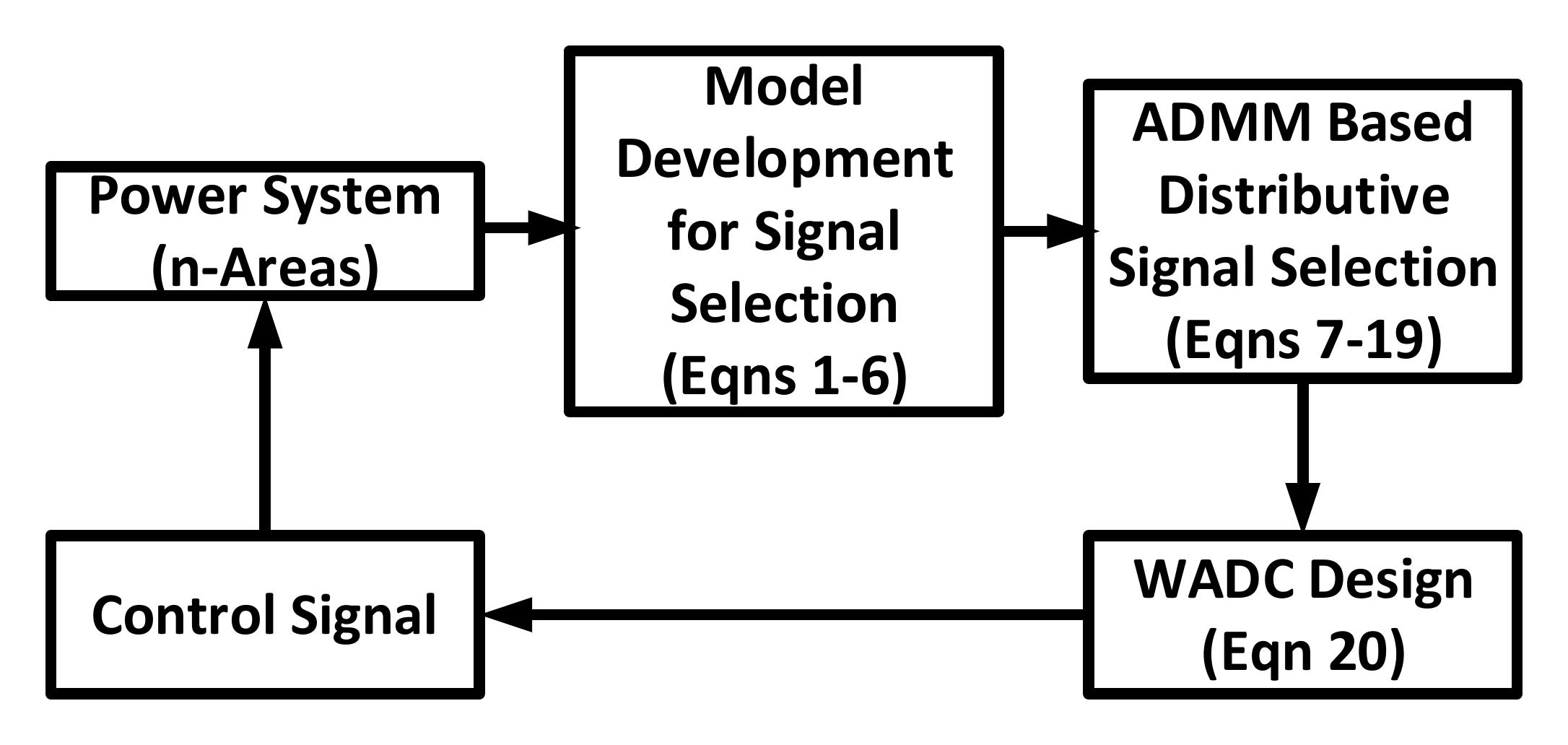}
\caption{Flowchart of overall methodology.}
\label{fig1aa}
\end{figure}

The proposed methodology involves three steps: a) Divide large-scale network into areas, b) Select the appropriate signal for control, and c) Design a wide-area damping controller. The proposed distributed architecture is shown in Fig. \ref{fig1}. The flowchart is shown in Fig. \ref{fig1aa}.

\vspace{-3mm}
\subsection{Dividing Large Scale Network into Areas}
\label{coher}
Most of the present techniques for wide-area control are based on linearizing the power system model at an operating point and then calculating the eigenvalues and vectors to identify the optimal wide-area control loop. However getting a linearized system for large-scale utility network is complex and impractical. Also as the operating condition of the grid changes, calculating the linear system model for every time-step is impractical. To overcome this modeling difficulty, in this paper, a measurement based identification of wide-area control loop is proposed. However analyzing measurements data using centralized data processing framework may not be possible due to data transfer bottlenecks, data volume, as well as the non-availability of communication infrastructure. Also, in reality, different utilities (areas) are combined to a larger power system, so for centralized processing, detailed information of other areas may be required. This can be a difficult task due to the nonavailability of the data at the central control center (global processor). To overcome this, the power system is divided into areas based on utility boundaries and each utility is connected to the other through tie-lines. If utility boundary information is not provided then the power system is divided into areas based on coherency grouping of generators such that each area has one local processor.

In this paper, the coherency grouping is used to divide the system into physical areas (utilities) at one operating condition to place the local processors as the test system has no information of physical areas. For the division of the network into areas, an online coherency grouping based on spectral clustering methodology considering the speed deviation of the generators is used. Let the speed deviation ($rated-actual$) data points $\Delta\omega_{1},\Delta\omega_{2},... \Delta\omega_{n}$ for a window length of $n$ are considered for clustering. Using these data points a similarity matrix $S$ is formulated, where $S_{ij}$ gives the relation between $\Delta\omega_{i}$ and $\Delta\omega_{j}$. The information from similarity matrix is used to group $\Delta\omega_{1},\Delta\omega_{2},... \Delta\omega_{n}$ into $k$ clusters. The similarity matrix is based on a Gaussian function represented as in \eqref{eqn1n}

\begin{eqnarray}
S_{ij}=e^{-\frac{\left\| \Delta\omega_{i}-\Delta\omega_{j}\right\|}{2 \sigma^2}}
\label{eqn1n}
\end{eqnarray}
where $\sigma$ is a scaling factor. Here  $S$ is dense and is of the order $n\times n$. The size of $S$ increases with an increase in the number of data points under consideration, but this slows the simulation speed. To increase the online coherency grouping speed, Nystrom method is used which uses sub-matrix of the dense matrix $\textbf{S}$ \cite{refsc}. The details regarding online coherency grouping methodology are discussed in \cite{refa11}. It is worth noting that, the coherency grouping changes as the system operating condition changes, however in this paper the objective of coherency grouping is to divide the system into physical areas for implementing the proposed algorithm. In the future work the dynamic grouping will be considered in the design process.

\subsubsection{Dividing the two-area system}
\begin{figure}[!h]
\centering
\includegraphics[width=0.475\textwidth]{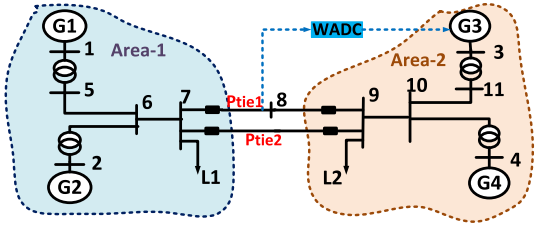}
\caption{Two area test system.}
\label{fig1k}
\end{figure}

\begin{figure}[!h]
\centering
\includegraphics[width=0.475\textwidth]{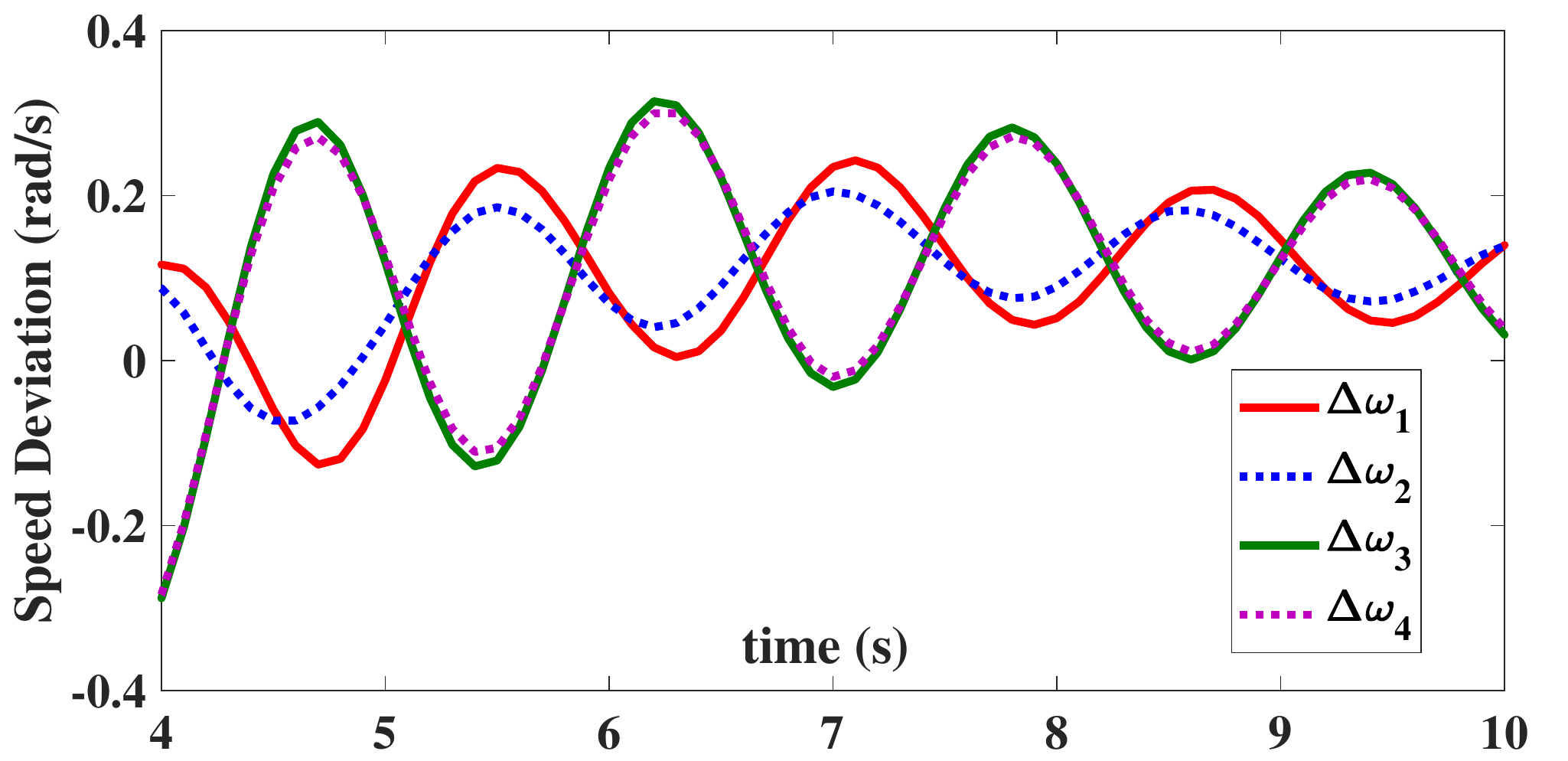}
\caption{Speed deviation of generators(two-area)}
\label{fig1z}
\end{figure}

For example, the coherency grouping algorithm is implemented on a two-area power system model which consists of four generators each with a capacity of 900 MVA as shown in Fig. \ref{fig1k}. The first step in implementing the proposed algorithm is to identify/divide the system into areas based on coherency grouping or based on the real physical geography of the larger power system. For this a  3-ph fault is created for a duration of 0.1 sec on bus-9. The speed deviation data obtained after the fault is then  analyzed to develop coherent groups of generators. The speed deviations of all generators are shown in Fig. \ref{fig1z}. From the analysis it can be found that generators 1, 2 are in one area and generators 3, 4 are in the other area (see Table \ref{table1}). From Fig. \ref{fig1z} it can be seen that the generators 1 and 2 are connected to generators 3 and 4 through two tie-lines between Bus-7 and Bus-9.

\begin{table}[h]
\renewcommand{\arraystretch}{1.3}
\centering
\caption{COHERENCY GROUPING OF GENERATORS}
\label{table1}
\begin{tabular}{*9c} 
\toprule
 Test System & Grouping\\
\hline
Two Area  & Group-1: 1,2\\
      & Group-2: 3,4\\
\hline
IEEE 39-BUS & Group-1: 4,5,6,7,9\\
    & Group-2: 1,8\\
     & Group-3: 2,3\\
     & Group-4: 10\\
\hline
\end{tabular}
\end{table}

\subsubsection{Dividing the IEEE-39 bus system}
\begin{figure}[!h]
\centering
\includegraphics[width=0.475\textwidth]{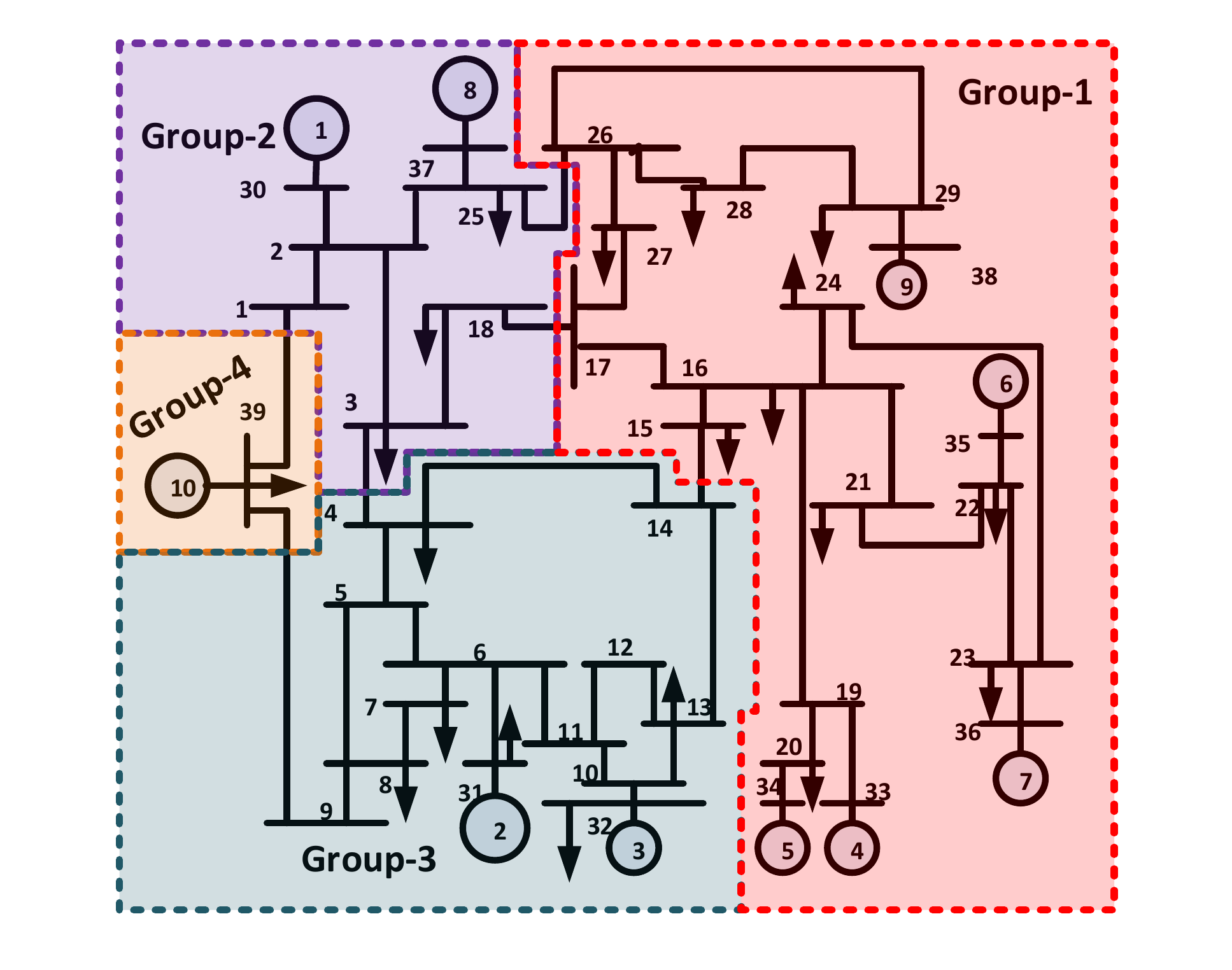}
\caption{IEEE-39 bus test system model.}
\label{fig2a}
\end{figure}

\begin{figure*}[!h]
\centering
\includegraphics[width=0.98\textwidth]{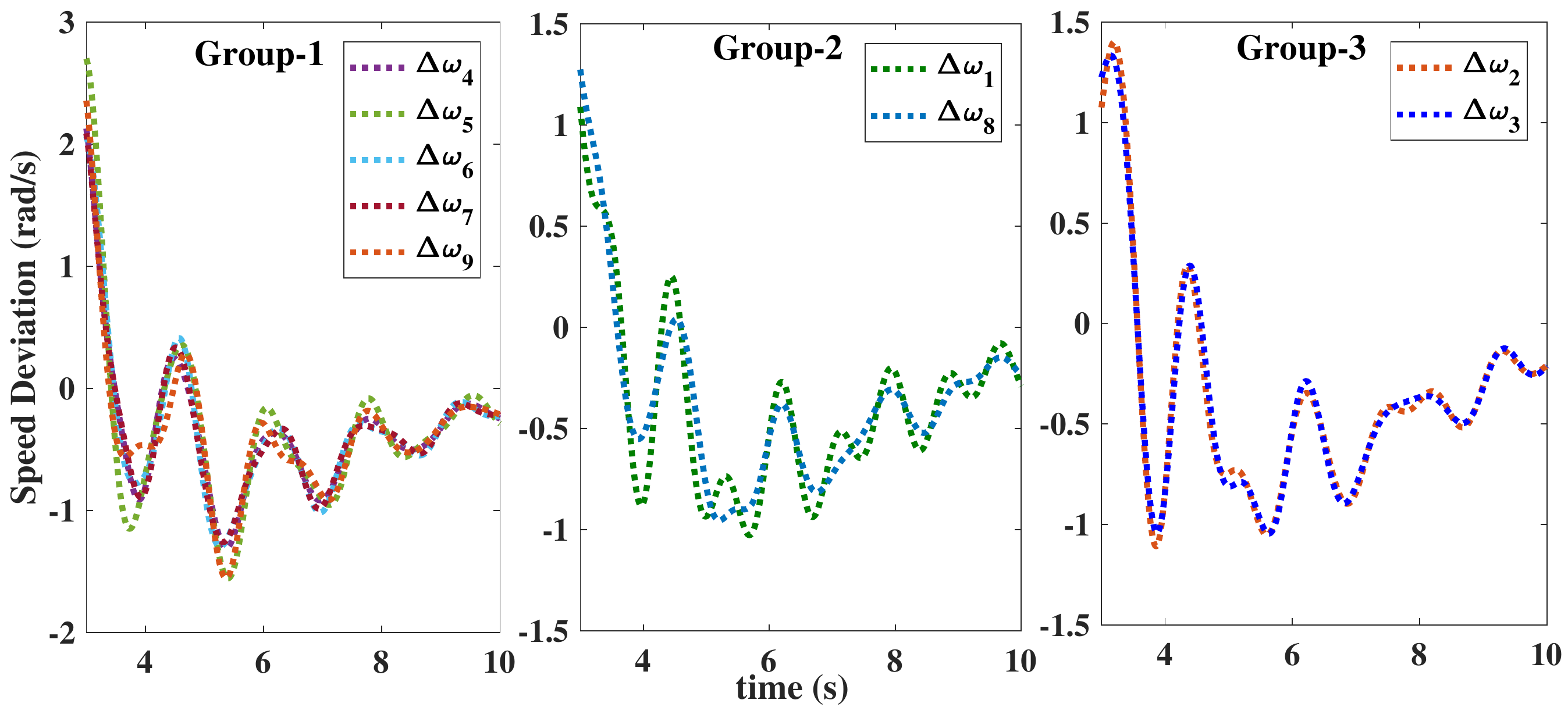}
\caption{Relative speeds of generators (39-bus)}
\label{fig2aa}
\end{figure*}

The algorithm scalability is tested on a real-life power grid, IEEE 39 bus system (Fig. \ref{fig2a}), which consists of 39 buses and 10 generators.  For grouping this power grid, a three-phase fault is created on Bus-14 at 2 sec for a duration of 0.1 sec. The speed deviations are analyzed to identify the coherent groups of generators. The groups of generators are as shown in Table. \ref{table1} and Fig. \ref{fig2aa}. The network is divided into four areas such that each coherent group of generators area in one area and are connected by tie-lines.

\subsection{Proposed Architecture for Signal Selection} \label{signalsel}
The proposed architecture for optimal wide-area signal selection involves the following steps: a) Model development, and b) ADMM based distributive signal selection.
\subsubsection{Model Development}
\begin{figure}[!h]
\centering
\includegraphics[width=0.475\textwidth,height=1.8in]{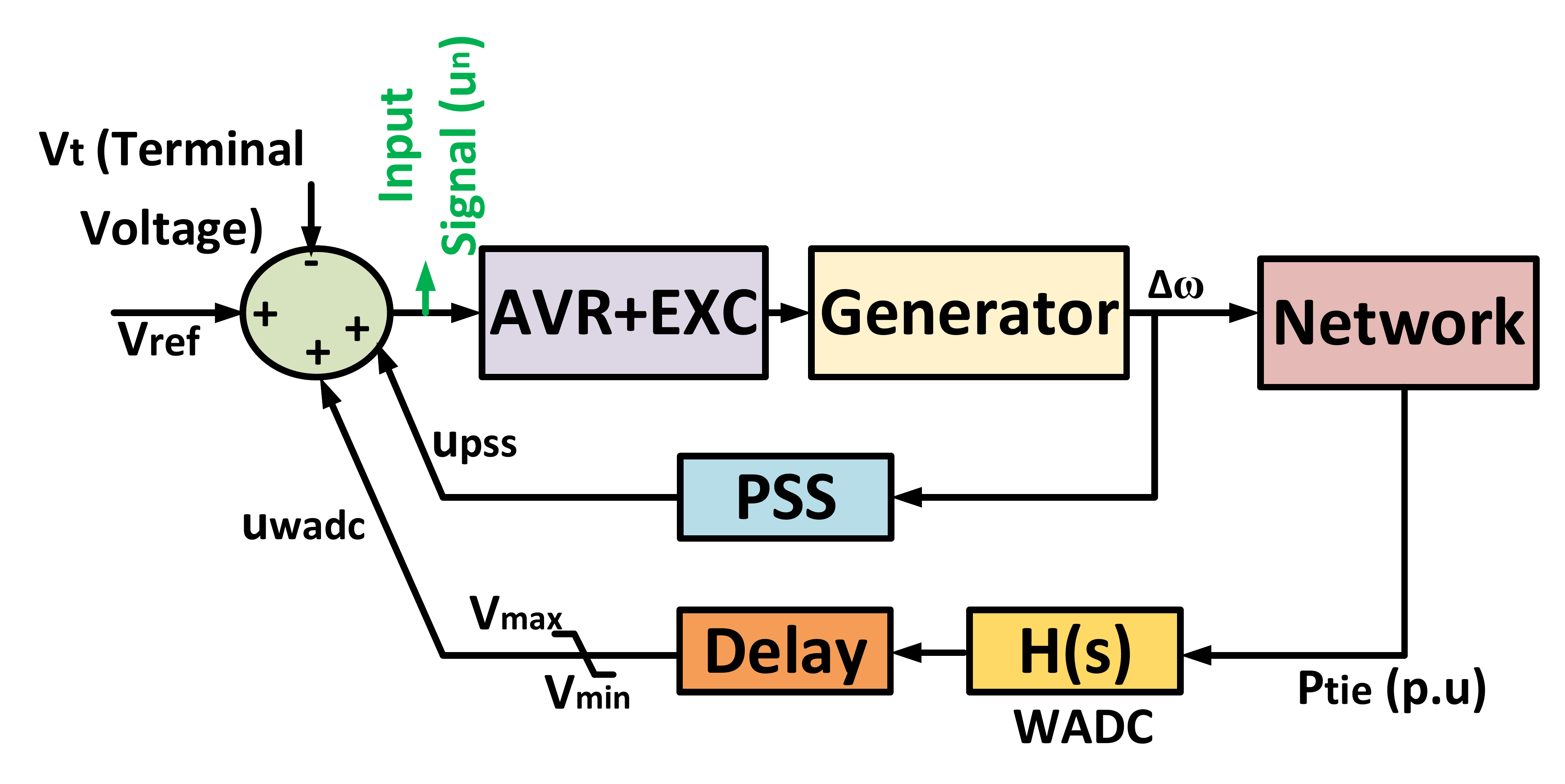}
\caption{Wide area control implementation.}
\label{fig4a}
\end{figure}

First, the power system is divided into areas. It is assumed that each area has a local processor which communicates with the global processor as a local processor of one area may not have complete information of other areas for determining optimal wide area control loop. The local processor estimates the local transfer functions, for this the generator exciter signal $u_n$ as shown in Fig. \ref{fig4a} is used. This means the output of the WADC is fed to the generator for damping inter-area oscillations. In the distributed algorithms it is required to identify the detection points which are common to multiple areas such that each local processor uses that information as output for transfer function estimation. This is also required for convergence of all areas to a common point. In the power grid, the tie-line information is common to multiple areas and local processors can use tie-line power flow to reach a consensus and for controller convergence. Since the inter-area oscillations (0.1 to 1 Hz) are between areas through tie-lines, the active power flow through the tie-lines capture the inter-area oscillations and hence an appropriate signal for wide area control. Thus a local transfer function is estimated with input signal ($u_n$) and tie-line power flow as output \cite{refa12}.

For example, if there are $m$ tie lines and $n$ generators in the system, then the MIMO transfer function of the power system can be written as
 \begin{eqnarray}
  \left[{\begin{array}{c}
   P_{1}(z) \\
   . \\
   . \\
   P_{m}(z) 
  \end{array} }\right] = \left[{\begin{array}{ccc}
   G_{11}(z) & \dots & G_{1n}(z)\\
   . & \dots & .\\
   . & \dots & .\\
   G_{m1}(z) & \dots & G_{mn}(z) 
  \end{array} }\right]\left[{\begin{array}{c}
   u_{1}(z)\\
   .\\
   .\\
   u_{n}(z)
  \end{array} }\right]
  \label{eqn1}
\end{eqnarray}
and generalized as 
\begin{equation}
    P(z)=G(z)U(z)
    \label{eqn2}
\end{equation}
where $u_n$  is the input signal (see Fig. \ref{fig4a}) and $P_m$ is the tie-line power flow. 
Based on the MIMO transfer function of the power system, \eqref{eqn1}-\eqref{eqn2} can be represented as $G_{mn}(z)$ \cite{refa13,refa14}

\begin{equation}
    G_{mn}(z)=\frac{P_m(z)}{u_n(z)}=\frac{b_0^h+b_1^hz^{-1}+\dots+b_k^hz^{-k}}{1+a_1z^{-1}+\dots+a_kz^{-k}}
    \label{eqn3}
\end{equation}

\begin{figure*}[!h]
\normalsize
\setcounter{MYtempeqncnt}{\value{equation}}
\begin{equation}
\label{glo1}
\left[\begin{array}{c}
  P_m(j)\\
  P_m(j-1)\\
  \vdots\\
  P_m(j-N+1)
  \end{array}\right]
  =\left[\begin{array}{cccccc}
    P_m(j-1) & .. & P_m(j-k) & u_n(j) & .. & u_n(j-k) \\
    P_m(j-2) & .. & P_m(j-k-1) & u_n(j-1) & .. & u_n(j-k-1)\\
    \vdots & \vdots & \vdots & \vdots & \vdots & \vdots\\
    P_m(j-N) & .. & P_m(j-N+1-k) & u_n(j-N+1) & .. & u_n(j-N+1-k)
   \end{array}\right]
 \left[\begin{array}{c}
  a_1\\
  \vdots\\
  a_k\\
  b_0^n\\
  b_1^n\\
  \vdots\\
  b_k^n
  \end{array}\right]
\end{equation}
\hrulefill
\end{figure*}
where $h$  is the element number in the matrix and $k$ is the order of transfer function. It can be seen from \eqref{eqn3} that, for capturing inter-area modes in the power system, the denominator coefficients need to be the same however numerator coefficients should be different. This gives information regarding inter-area modes in the system as seen from different inputs but their corresponding residues differ. For $j$  samples and $N$  observation window length, \eqref{eqn3} can be rewritten as shown in \eqref{glo1}. The least squares technique is used to estimate \eqref{eqn3} in discrete time-domain. $G(z)$ in \eqref{eqn2} is formulated after estimating the entire system with different inputs and outputs. Now \eqref{eqn3} can be transformed into partial fraction expansion and can be written as \eqref{eqn4}.
\begin{equation}
\label{eqn4}
\begin{aligned}
G_{mn}(z)=\frac{P_m(z)}{u_n(z)}=\frac{R_{(mn)k}}{z-p_{(mn)k}}+\cdots+\frac{R_{(mn)2}}{z-p_{(mn)2}}+\\
\frac{R_{(mn)1}}{z-p_{(mn)1}}+k(z)
\end{aligned}
\end{equation}
where $k(z)$  is a polynomial in $z$ and $R_{(mn)k}$ is the residue of  corresponding to the pole $p_{(mn)k}$. The residue  $R_{(mn)k}$ provides information about how the mode $p_{(mn)k}$ is affected by input $u_n$ and how observable is it from $P_m$. This shows that residue is a measure of joint controllability and observability index, where larger the value of residue, the stronger is the optimal control loop. However, solving \eqref{eqn1} using centralized techniques may not be possible because of computational burden, data volume etc. For example, if there are 20 tie-lines and 200 generators, using a centralized method the processor must solve 4000 transfer functions which may not be feasible in real-time. On the other hand, the power system is divided into areas either based on geographical location (utilities) or coherency grouping and each area has its own local processor which reports to global processor. Thus local processor of one area (utility) may not have complete information of other areas for wide area control. In the proposed distributed architecture, a distributed consenses based approach using ADMM can be used to overcome this problem by solving \eqref{eqn1} in a distributed architecture. In this approach, using Lagrange multipliers method a black-box transfer function model is estimated locally for each area and then a local processor is used to share information with the global processor so that a global transfer function of the power system can be estimated. The eigenvalue and corresponding residue information obtained from the global transfer function is used in identifying a wide-area control loop and controller design.
\subsubsection{ADMM Based Distributive Signal Selection}
Large power system comprises of different areas, so \eqref{eqn1} can be divided into parts and then solved using ADMM. The distributed algorithm is proposed here using an example for simplicity and ease of understanding. Suppose there are four generators ($n=4$) divided into two areas such that generators 1 and 2 are in one area and generators 3 and 4 are in the other area, both these areas are connected by two tie-lines ($m=2$). Then \eqref{eqn1} can be written as follows:
\begin{eqnarray}
  \left[{\begin{array}{c}
   P_{1}(z) \\
   P_{2}(z) 
  \end{array} }\right] = \left[{\begin{array}{cccc}
  G_{11}(z) & \cdots & G_{14}(z)\\
  G_{21}(z) & \cdots & G_{24}(z)
  \end{array} }\right]\left[{\begin{array}{c}
   u_{1}(z)\\
   u_{2}(z)\\
   u_{3}(z)\\
   u_{4}(z)
  \end{array} }\right]
  \label{eqn5}
\end{eqnarray}
The centralized equation \eqref{eqn5} can be distributed and reformulated as follows:\\
\textbf{Step-1:} Divide the above MIMO transfer function into two areas:
\begin{itemize}
  \item \textit{For Area-1}
\end{itemize}
\begin{eqnarray}
  \left[{\begin{array}{c}
   P_{1}(z) \\
   P_{2}(z) 
  \end{array} }\right] = \left[{\begin{array}{cc}
  G_{11}(z) & G_{12}(z)\\
  G_{21}(z) & G_{22}(z)
  \end{array} }\right]\left[{\begin{array}{c}
   u_{1}(z)\\
   u_{2}(z)
  \end{array} }\right]
  \label{eqn6}
\end{eqnarray}
Further $G_{11}(z)$, $G_{12}(z)$, $G_{21}(z)$, and $G_{22}(z)$ can be written as follows:
\begin{align}
\begin{split}
   G_{11}(z)=\frac{P_1(z)}{u_1(z)}=\frac{b_0^1+b_1^1z^{-1}+\cdots+b_k^1z^{-k}}{1+a_1z^{-1}+\cdots+a_kz^{-k}} \\
    G_{12}(z)=\frac{P_1(z)}{u_2(z)}=\frac{b_0^2+b_1^2z^{-1}+\cdots+b_k^2z^{-k}}{1+a_1z^{-1}+\cdots+a_kz^{-k}} \\
    G_{21}(z)=\frac{P_2(z)}{u_1(z)}=\frac{b_0^3+b_1^3z^{-1}+\cdots+b_k^3z^{-k}}{1+a_1z^{-1}+\cdots+a_kz^{-k}} \\
    G_{22}(z)=\frac{P_2(z)}{u_2(z)}=\frac{b_0^4+b_1^4z^{-1}+\cdots+b_k^4z^{-k}}{1+a_1z^{-1}+\cdots+a_kz^{-k}}
\end{split}
  \label{eqn7}   
\end{align}
where $b_0^h$, $b_1^h$, $\cdots$, $b_k^h$ are numerator coefficients which are different for each transfer function, $a_1$, $a_2$, $\cdots$, $a_k$ are the denominator coefficients which are equal for all the transfer functions in the power system and $h=1,\cdots,4$.
\begin{itemize}
  \item \textit{For Area-2}
\end{itemize}
\begin{eqnarray}
  \left[{\begin{array}{c}
   P_{1}(z) \\
   P_{2}(z) 
  \end{array} }\right] = \left[{\begin{array}{cc}
  G_{13}(z) & G_{14}(z)\\
  G_{23}(z) & G_{24}(z)
  \end{array} }\right]\left[{\begin{array}{c}
   u_{3}(z)\\
   u_{4}(z)
  \end{array} }\right]
  \label{eqn8}
\end{eqnarray}
Further $G_{13}(z)$, $G_{14}(z)$, $G_{23}(z)$, and $G_{24}(z)$ can be written as follows:
\begin{align}
\begin{split}
   G_{13}(z)=\frac{P_1(z)}{u_3(z)}=\frac{b_0^5+b_1^5z^{-1}+\cdots+b_k^5z^{-k}}{1+a_1z^{-1}+\cdots+a_kz^{-k}} \\
    G_{14}(z)=\frac{P_1(z)}{u_4(z)}=\frac{b_0^6+b_1^6z^{-1}+\cdots+b_k^6z^{-k}}{1+a_1z^{-1}+\cdots+a_kz^{-k}} \\
    G_{23}(z)=\frac{P_2(z)}{u_3(z)}=\frac{b_0^7+b_1^7z^{-1}+\cdots+b_k^7z^{-k}}{1+a_1z^{-1}+\cdots+a_kz^{-k}} \\
    G_{24}(z)=\frac{P_2(z)}{u_4(z)}=\frac{b_0^8+b_1^8z^{-1}+\cdots+b_k^8z^{-k}}{1+a_1z^{-1}+\cdots+a_kz^{-k}}
\end{split}
  \label{eqn9}   
\end{align}
where $b_0^h$, $b_1^h$, $\cdots$, $b_k^h$ are numerator coefficients which are different for each transfer function, $a_1$, $a_2$, $\cdots$, $a_k$ are the denominator coefficients which are equal for all the transfer functions in the power system and $h=5,\cdots,8$.\\
\textbf{Step-2:} Writing the \eqref{eqn7} and \eqref{eqn9} in least squares format
\begin{itemize}
  \item \textit{For Area-1}
\end{itemize}
\begin{eqnarray}
\begin{split}
   \left[{\begin{array}{cc}
  L_1 & M_1
  \end{array} }\right]\left[{\begin{array}{c}
  a^1\\
  b^1
  \end{array} }\right] = \left[B_{11}\right]\\
  \left[{\begin{array}{cc}
  L_1 & M_2
  \end{array} }\right]\left[{\begin{array}{c}
  a^2\\
  b^2
  \end{array} }\right] = \left[B_{12}\right]\\
  \left[{\begin{array}{cc}
  L_2 & M_1
  \end{array} }\right]\left[{\begin{array}{c}
  a^3\\
  b^3
  \end{array} }\right] = \left[B_{21}\right]\\
  \left[{\begin{array}{cc}
  L_2 & M_2
  \end{array} }\right]\left[{\begin{array}{c}
  a^4\\
  b^4
  \end{array} }\right] = \left[B_{22}\right]
\end{split}
  \label{eqn10}   
\end{eqnarray}
\begin{itemize}
  \item \textit{For Area-2}
\end{itemize}
\begin{eqnarray}
\begin{split}
   \left[{\begin{array}{cc}
  L_1 & M_3
  \end{array} }\right]\left[{\begin{array}{c}
  a^5\\
  b^5
  \end{array} }\right] = \left[B_{13}\right]\\
  \left[{\begin{array}{cc}
  L_1 & M_4
  \end{array} }\right]\left[{\begin{array}{c}
  a^6\\
  b^6
  \end{array} }\right] = \left[B_{14}\right]\\
  \left[{\begin{array}{cc}
  L_2 & M_3
  \end{array} }\right]\left[{\begin{array}{c}
  a^7\\
  b^7
  \end{array} }\right] = \left[B_{23}\right]\\
  \left[{\begin{array}{cc}
  L_2 & M_4
  \end{array} }\right]\left[{\begin{array}{c}
  a^8\\
  b^8
  \end{array} }\right] = \left[B_{24}\right]
\end{split}
  \label{eqn11}   
\end{eqnarray}
where $a$ is vector of denominator coefficients, $b$  is vector of numerator coefficients, $L$  is matrix of previous samples of $P_m$, and $M_n$ is matrix of current and previous samples of $u_n$.\\
\textbf{Step-3:} The objective here is to make $a^1=a^2=\cdots=a^8=z$   for a global consensus problem so that with the given initial conditions numerator and denominator coefficients can be calculated iteratively until objective is achieved.
\begin{itemize}
  \item \textit{For Area-1 and Area-2 (Calculating $a$ using $b$)}
\end{itemize}
\begin{align}
\begin{split}
   \left[L_1\right]\left[a^1\right]=\left[B_{11}\right]-\left[M_1\right]\left[b^1\right]\\
    \left[L_1\right]\left[a^2\right]=\left[B_{12}\right]-\left[M_2\right]\left[b^2\right]\\
   \left[L_2\right]\left[a^3\right]=\left[B_{21}\right]-\left[M_1\right]\left[b^3\right]\\
   \left[L_2\right]\left[a^4\right]=\left[B_{22}\right]-\left[M_2\right]\left[b^4\right]\\
   \left[L_1\right]\left[a^5\right]=\left[B_{13}\right]-\left[M_3\right]\left[b^5\right]\\
   \left[L_1\right]\left[a^6\right]=\left[B_{14}\right]-\left[M_4\right]\left[b^6\right]\\
  \left[L_2\right]\left[a^7\right]=\left[B_{23}\right]-\left[M_3\right]\left[b^7\right]\\
  \left[L_2\right]\left[a^8\right]=\left[B_{24}\right]-\left[M_4\right]\left[b^8\right]
\end{split}
  \label{eqn12}   
\end{align}
\begin{itemize}
  \item \textit{For Area-1 and Area-2 (Calculating $b$ using $a$)}
\end{itemize}
\begin{align}
\begin{split}
   \left[M_1\right]\left[b^1\right]=\left[B_{11}\right]-\left[L_1\right]\left[a^1\right]\\
    \left[M_2\right]\left[b^2\right]=\left[B_{12}\right]-\left[L_1\right]\left[a^2\right]\\
   \left[M_1\right]\left[b^3\right]=\left[B_{21}\right]-\left[L_2\right]\left[a^3\right]\\
   \left[M_2\right]\left[b^4\right]=\left[B_{22}\right]-\left[L_2\right]\left[a^4\right]\\
   \left[M_3\right]\left[b^5\right]=\left[B_{13}\right]-\left[L_1\right]\left[a^5\right]\\
   \left[M_4\right]\left[b^6\right]=\left[B_{14}\right]-\left[L_1\right]\left[a^6\right]\\
  \left[M_3\right]\left[b^7\right]=\left[B_{23}\right]-\left[L_2\right]\left[a^7\right]\\
  \left[M_4\right]\left[b^8\right]=\left[B_{24}\right]-\left[L_2\right]\left[a^8\right]
\end{split}
  \label{eqn13}   
\end{align}
More generically \eqref{eqn12} and \eqref{eqn13} can be rewritten as follows:
\begin{equation}
    \label{eqn14}
     \left[L^q\right]\left[a^q\right]=\left[B^q\right]-\left[M^q\right]\left[b^q\right]
\end{equation}
\begin{equation}
    \label{eqn15}
    \left[M^q\right]\left[b^q\right]=\left[B^q\right]-\left[L^q\right]\left[a^q\right]
\end{equation}
Where $q$ is area number.\\
\textbf{Step-4:} Global consensus optimization problem is formulated using equations \eqref{eqn14} and \eqref{eqn15} as follows:
\begin{equation}
\begin{aligned}
& \underset{a^1,\cdots,a^q,z}{\text{min}}
& & \sum_{q=1}^{2}\frac{1}{2}\left\|\left[L^q\right]\left[a^q\right]-\left[B^q\right]+\left[M^q\right]\left[b^q\right]\right\|^2 \\
& \text{subject to}
& & a^q-z=0, for\ q=1,2
\end{aligned}
\label{eqn16}
\end{equation}
 $z$ is the global consensus solution, that is obtained when the local estimates of all local processors denoted by $a^q$, $q=1,2$ reach the same value.

It can be seen that the ADMM is an estimation method uses Lagrange multiplier approach in an iterative distributed methodology. The augmented Lagrange is computed as follows \cite{refa15}
\begin{equation}
\begin{aligned}
L_{\rho}=& \sum_{q=1}^{2}\frac{1}{2}\left\|\left[L^q\right]\left[a^q\right]-\left[B^q\right]+\left[M^q\right]\left[b^q\right]\right\|+\\
& w_q^T(a^q-z)+\frac{\rho}{2}\left\|a^q-z\right\|^2
\end{aligned}
\label{eqn17}
\end{equation}
where $a$ and $z$  are the vectors of the primal variables, $w$  is the vector of the dual variables or the Lagrange multipliers associated with \eqref{eqn16}, and $\rho>0$ denotes a penalty factor. ADMM implementation to solve distributed MIMO system is shown in Algorithm \ref{algx1}. 
\begin{algorithm}
\caption{ADMM-Algorithm}
\label{algx1}
\begin{algorithmic}[1] 
\STATE Each local processor $(q)$ initializes $a_0^q$, $b_0^q$   using \eqref{eqn10} and \eqref{eqn11}. $z_0$  and $w_0^q$  are also initialized
\STATE At iteration j:
\STATE Local processors updates $a^q$ as $a_{j+1}^q=\underset{a^q}{\operatorname{argmin}}\ L_{\rho}$
\STATE Local processor transmits $a_{j+1}^q$ to the global processor
\STATE Global processor calculates $z_{j+1}=\frac{1}{2}\sum_{q=1}^{2}a_{j+1}^q$
\STATE Global processor transmits $z_{j+1}$  to all local processors.
\STATE Local processor updates $w^q$ as $w_{j+1}^q=w_j^q+\rho(a_{j+1}^q-z^{k+1})$
\STATE Local processor updates $b_{j+1}^q$  using \eqref{eqn13}
\end{algorithmic} 
\end{algorithm}
The advantages of using the Lagrange multiplier method is as follows. First the constraints can be included in the form of a norm. Second as we have only the equality constraints the approach is very feasible. Third these methods can significantly reduce the computational time for optimization. 
\subsection{Wide Area Damping Controller Design}
To provide increased damping of inter-area oscillations supplementary control is required. The supplementary control is to be applied to the controllable generator and it should work in parallel with other local controls of the generator. This supplementary control is called wide-area damping control. The input-output signal selection methodology for wide area damping controller is as discussed in Section-\ref{signalsel}. In the literature various wide area control are reported, however in this paper wide area damping controller design based on residue as reported in \cite{refa1,refa17} is adapted since the major focus of this paper is to study distributed approach for signal selection.

\begin{figure}[!h]
\centering
\includegraphics[height=1.0in]{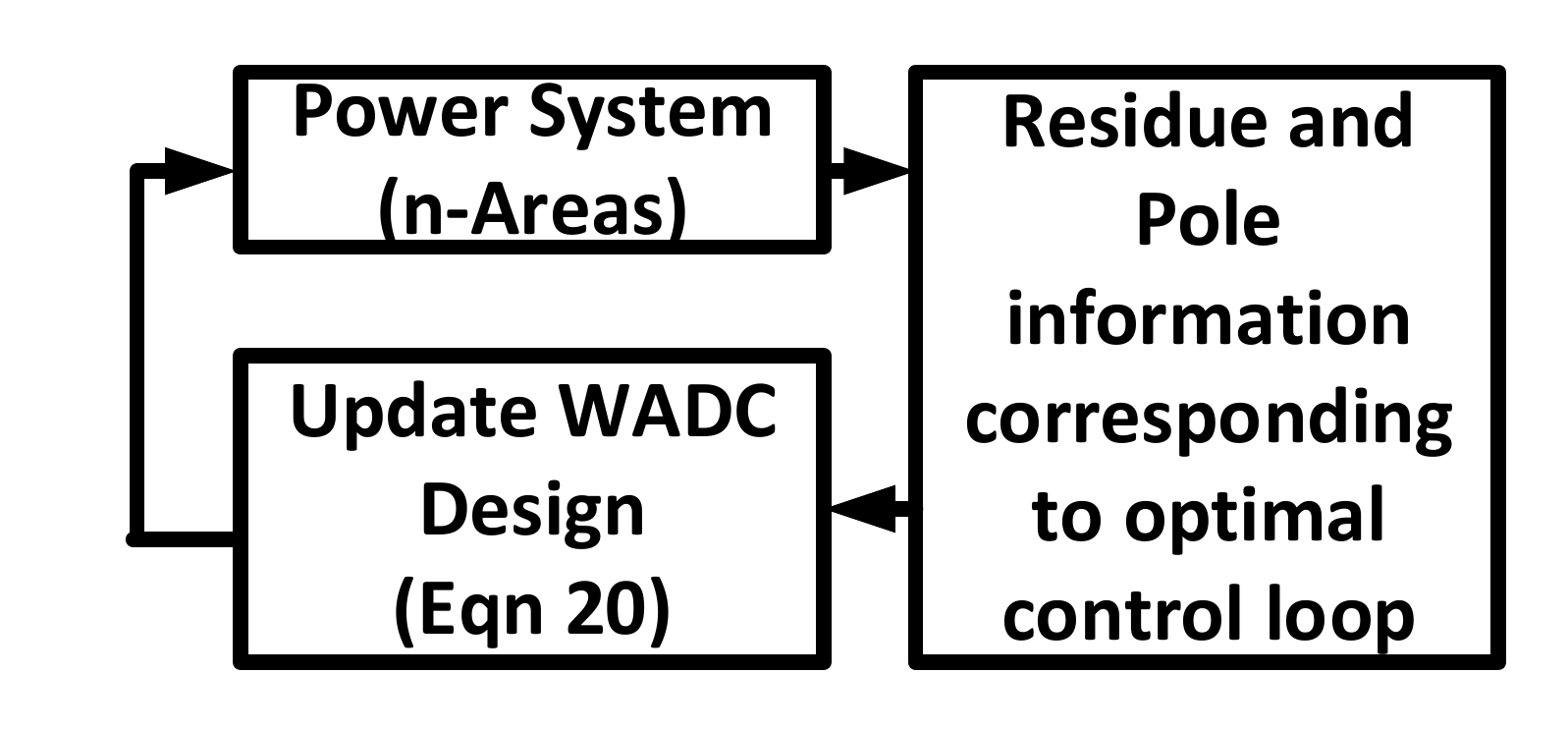}
\caption{Flowchart for WADC design.}
\label{figw}
\end{figure}

Fig. \ref{fig4a} shows the block diagram of the closed loop system with PSS and wide area controller $H(s)$. $H(s)$ is represented as in \eqref{eqn19}.
\begin{equation}
    H(s)=K_{WADC}\frac{sT_w}{1+sT_w}\left[\frac{1+sT_{lead}}{1+sT_{lag}}\right]^m
    \label{eqn19}
\end{equation}
where $K_{WADC}$ is the wide area controller gain,  $T_w$ is the washout time constant (usually 5 - 10 sec), $T_{lag}$  and $T_{lead}$  are the lead and lag time constant respectively, and $m$  is the number of compensating blocks. Fig. \ref{figw} shows the flowchart for WADC design. The methodology of designing parameters for WADC is discussed in \cite{refa1,refa17} so further details are not presented here.
\vspace{-4mm}

\section{Experimental Setup for Implementing the Proposed Signal Selection Method}
\begin{figure}[!h]
\centering
\includegraphics[width=0.5\textwidth,height=5in]{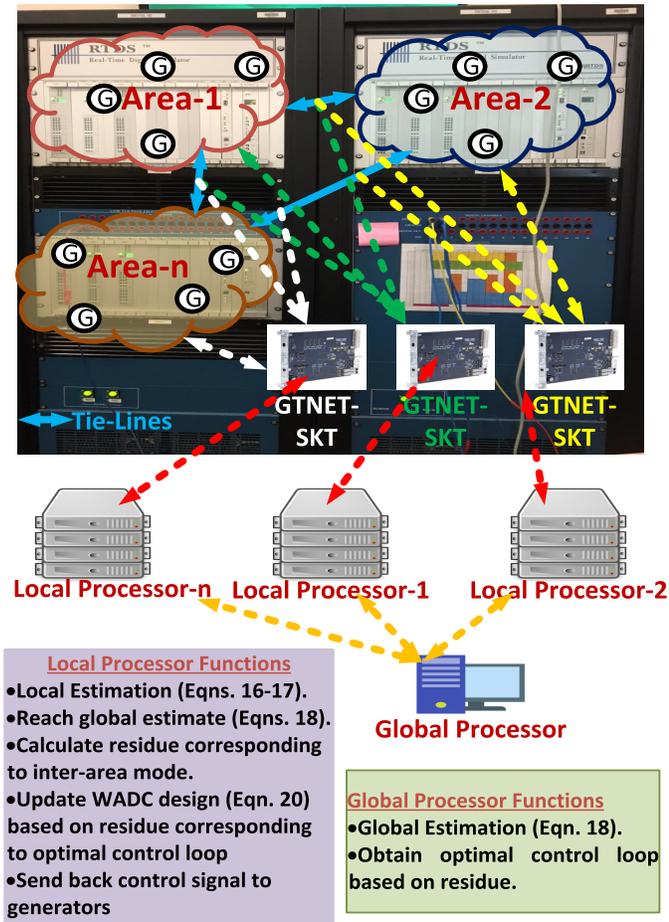}
\caption{Experimental test bed.}
\label{figaa}
\end{figure}

The proposed algorithm for signal selection and damping controller is implemented using RTDS/RSCAD and MATLAB co-simulation platform \cite{refa16}. The power system is modeled and simulated in RTDS/RSCAD whereas MATLAB sessions act as local and global processors. Both the MATLAB and RTDS communicate with each other using the GTNET-SKT hardware interface. Each area sends the required generator input $(u_n)$ and tie-line power flow data to the local processors through GTNET-SKT connection. Local processors will process the data,  estimate a local transfer function, and then communicates with the global processor to obtain a consensus-based global transfer function. Based on the global transfer function the wide area control loop and design of WADC are updated. Using the updated WADC design the local areas will send control actions to the generators. The simulation time step in RTDS in 50$\mu s$ (20000 samples/sec) but for the small signal stability analysis simulation time step of 20ms (50 samples/sec) is sufficient to preserve inter-area modes (0.1Hz to 1Hz). To reduce the data processing and computational time the data is down-sampled from 20000/sec to 50 samples/sec such that inter-area modes are preserved for small signal stability analysis. The experimental test bed is as shown in Fig. \ref{figaa}.

The local processors share information with the global processor through TCP/IP connection. For realizing this, different MATLAB sessions which act as local and global processors are assigned an IP address for sharing information. In every control architecture which involves communication, there will be inherent time delay. The time delay for wide-area control applications is of 100ms, so while designing the control architecture a time delay of 150 to 200ms should be considered \cite{del1}. To realize this feature in the control action, a 200ms time delay is added before sending the output of the controller to the generators. The communication protocols required for secure, industrial grade, standardized, scalable and distributed data communication infrastructure to support wide-area control in North America is recommended by the North American Synchro-Phasor Initiative through the introduction
of a NASPI network (NASPInet) \cite{naspi}.

\section{Implementation Test Results}

The proposed algorithm is initially implemented on two-area power system model which consists of four generators each 900 MVA. Then to further validate the algorithm on a larger system, IEEE 39 bus system which consists of 39 buses, 10 generators are used. The first step in implementing the algorithm is to identify/divide the system into areas based on coherency grouping or based on the real physical geography of the larger power system. Using the online coherency grouping technique two-area (Fig. \ref{fig1k}) and IEEE 39 bus system (Fig. \ref{fig2a}) are divided into groups as shown in Table. \ref{table1}.

\subsection{Implementation test results using two-area system}
\begin{figure}[!h]
\centering
\includegraphics[width=0.475\textwidth]{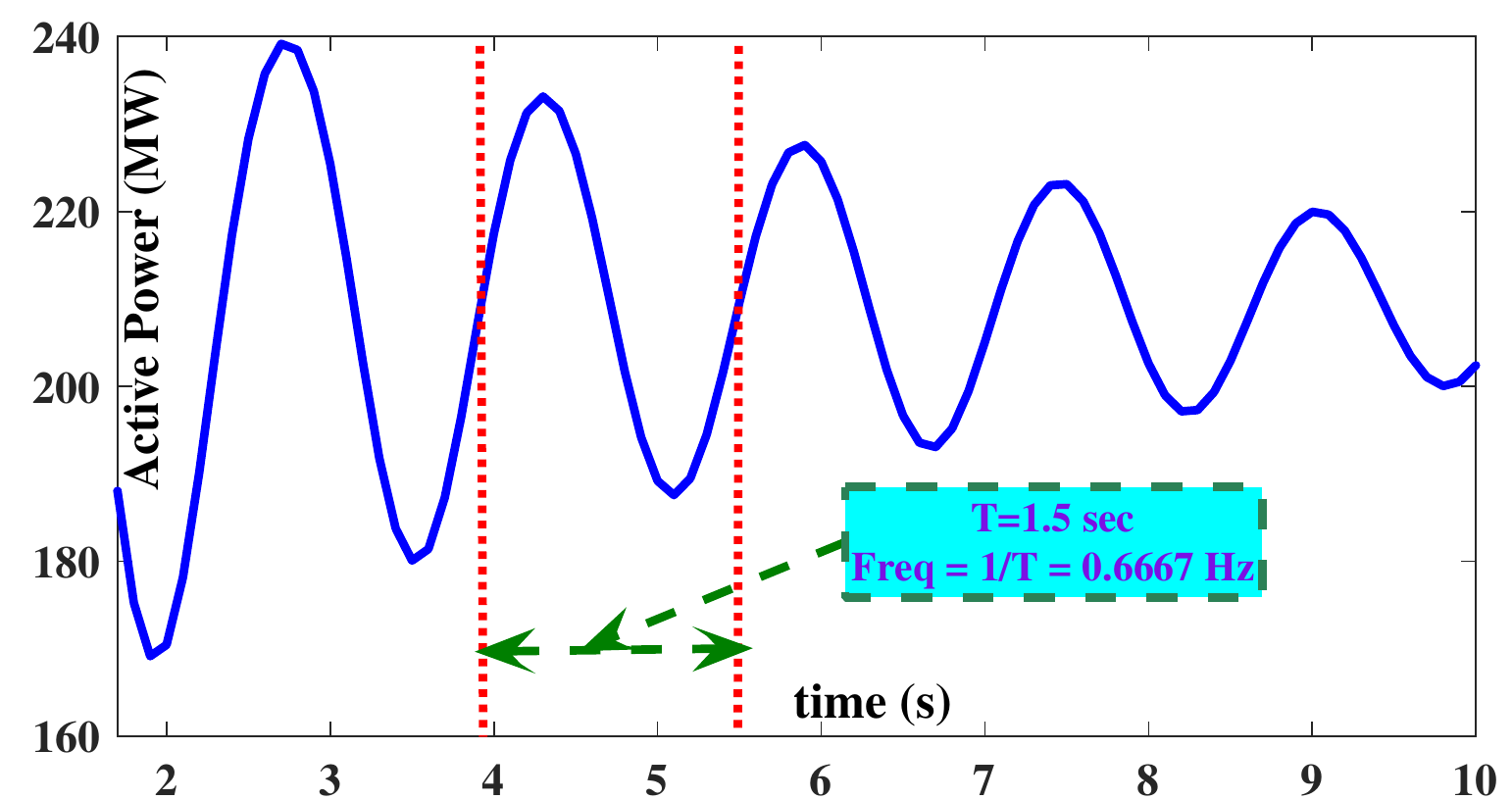}
\caption{Tie-line power flow (Bus7-Bus8-Bus9).}
\label{fig6aa}
\end{figure}

In this case, two-area system as shown in Fig. \ref{fig1k} is used. With an input signal as shown in Fig. \ref{fig4a} and tie line power flow as the output, MIMO transfer function is estimated locally for each area, and then local processor communicates with the global processor to arrive at a global solution. To validate the algorithm during an inter-area oscillation, a fault is created on bus 8 at 0.25sec and cleared at 0.45sec.
In this case residue analysis is performed by solving a global consensus problem. With estimation based on global consensus it is found that the frequency of oscillation is 0.6667 Hz which is in consensus with the tie-line power oscillation frequency as shown in Fig. \ref{fig6aa}. The frequency and residue (magnitude and angle) information give the optimal wide area control loop and this information is also critical for damping controller design. From Table. \ref{table1aa} it can be seen that the control loop between $P_{tie1}$ and Gen-3 has the highest value of residue for the inter-area mode of 0.6548 Hz. Based on this it can be concluded that Generator-3 is most controllable and tie-line (Bus7-Bus8-Bus9) power flow is the most observable signal. This means input to WADC should be the tie-line power flow and output should be fed to Generator-3.
To study the effectiveness of the proposed algorithm, the performance is tested with the presence of different generator controllers like:
\begin{itemize}
	\item With exciter
	\item With exciter and PSS
	\item With exciter and WADC
	\item With exciter, PSS, and WADC 
\end {itemize}

\begin{table}[!h]
\renewcommand{\arraystretch}{1.3}
\centering
\caption{RESIDUE ANALYSIS FOR SIGNAL SELECTION \\ (TWO-AREA SYSTEM)}
\label{table1aa}
\begin{tabular}{*9c} 
\toprule
  & \multicolumn{2}{c}{Residue} & \multicolumn{2}{c}{Frequency (Hz)}\\
\hline
 & $P_{tie1}$ & $P_{tie2}$ & $P_{tie1}$ & $P_{tie2}$\\
\hline
Gen-1 & 1.4649 & 1.3823 & 0.6563 & 0.6562\\
\hline
Gen-2 & 0.7893 & 0.7463 & 0.6572 & 0.6571\\
\hline
Gen-3 & 14.4958 & 13.3761 & 0.6548 & 0.6548\\
\hline
Gen-4 & 4.16 & 3.9402 & 0.6577 & 0.6578\\
\hline
\end{tabular}
\end{table}

\begin{figure}[!h]
\centering
\includegraphics[width=0.475\textwidth]{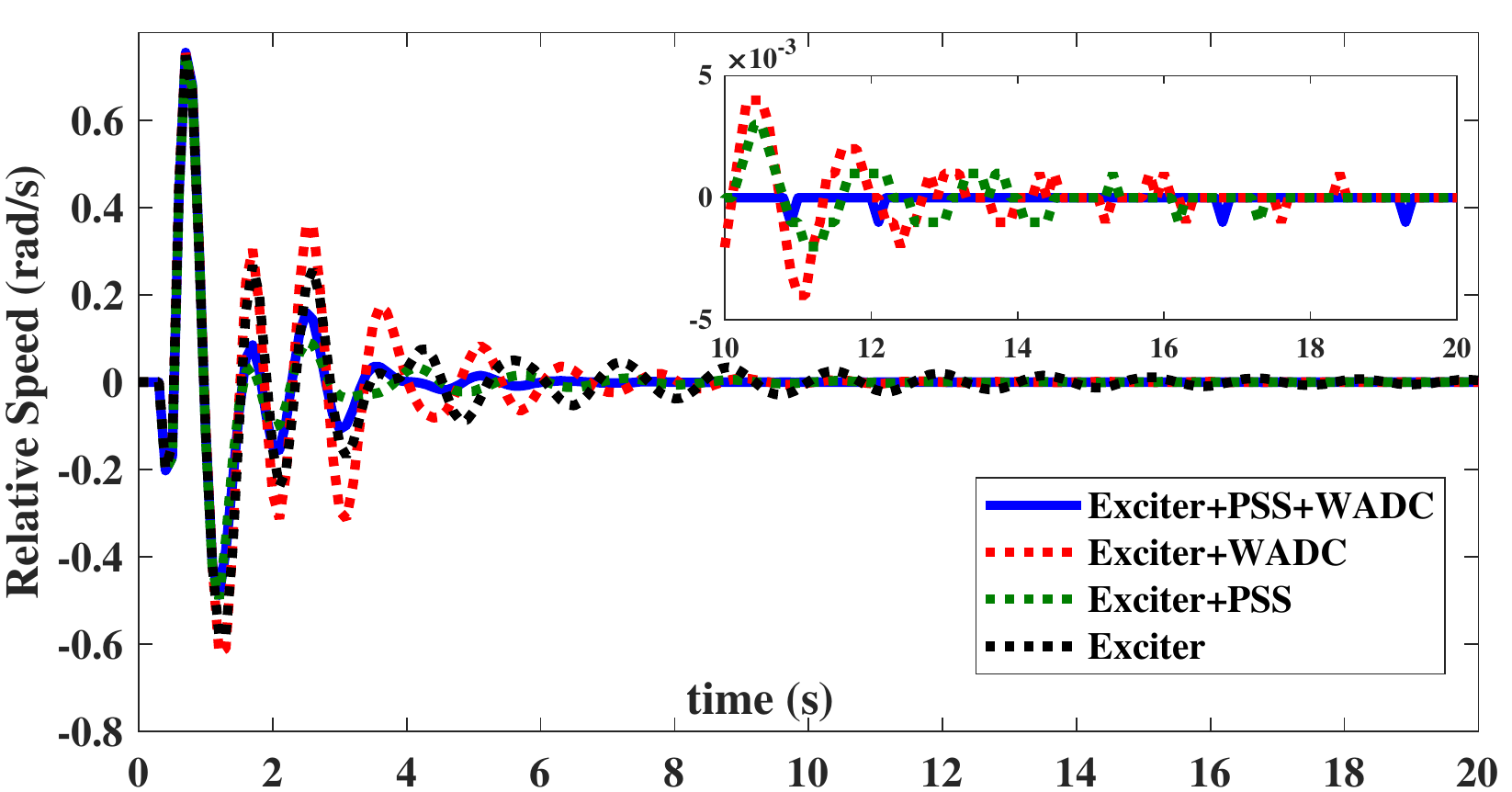}
\caption{Relative speed between generator-1 and generator-2.}
\label{fig7aa}
\end{figure}

\begin{figure}[!h]
\centering
\includegraphics[width=0.475\textwidth]{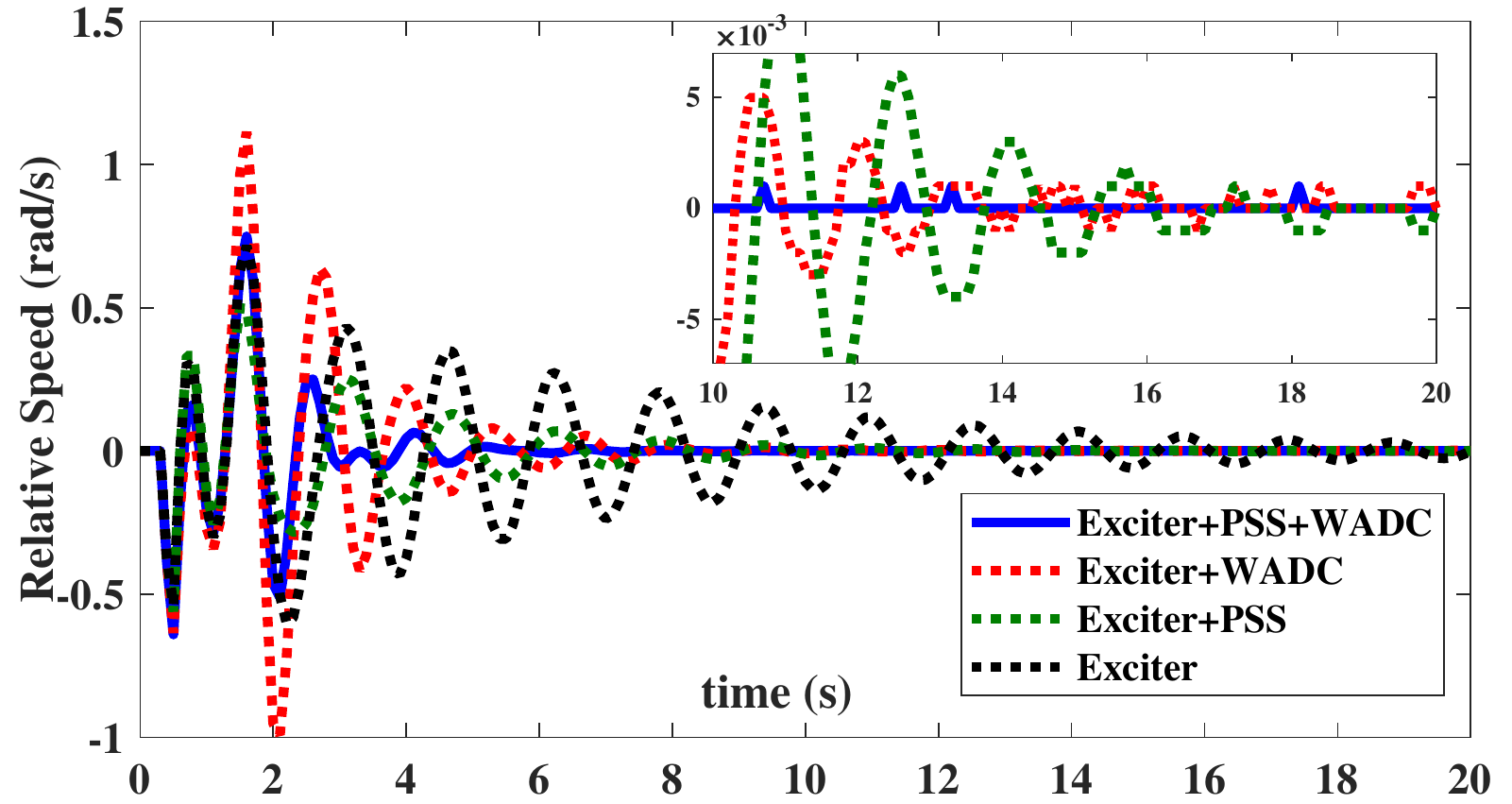}
\caption{Relative speed between generator-3 and generator-2.}
\label{fig8aa}
\end{figure}

\begin{figure}[!h]
\centering
\includegraphics[width=0.475\textwidth]{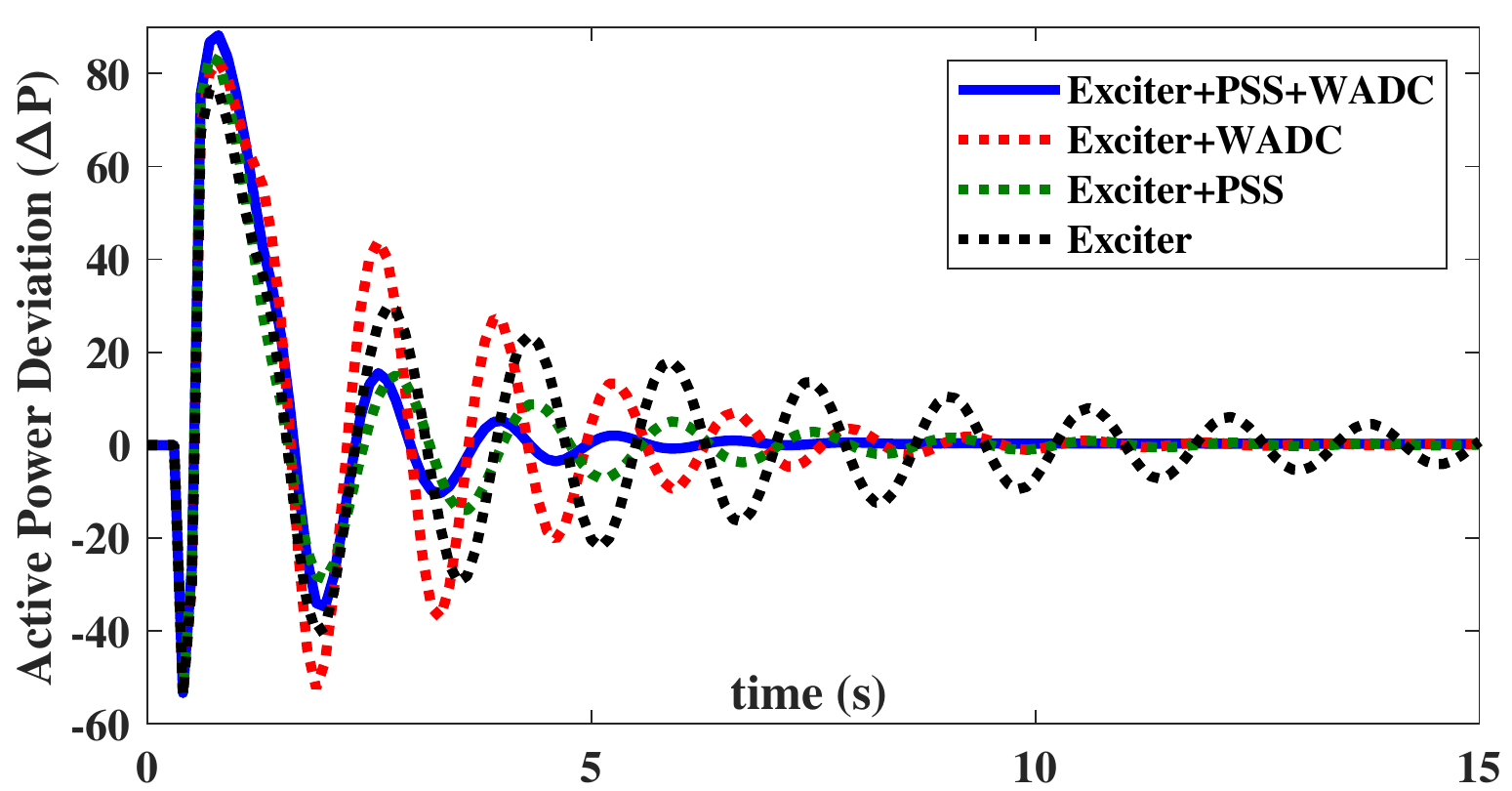}
\caption{Active power deviation through tie-line (Bus7-Bus8-Bus9).}
\label{fig9aa}
\end{figure}

\begin{figure}[!h]
\centering
\includegraphics[width=0.475\textwidth]{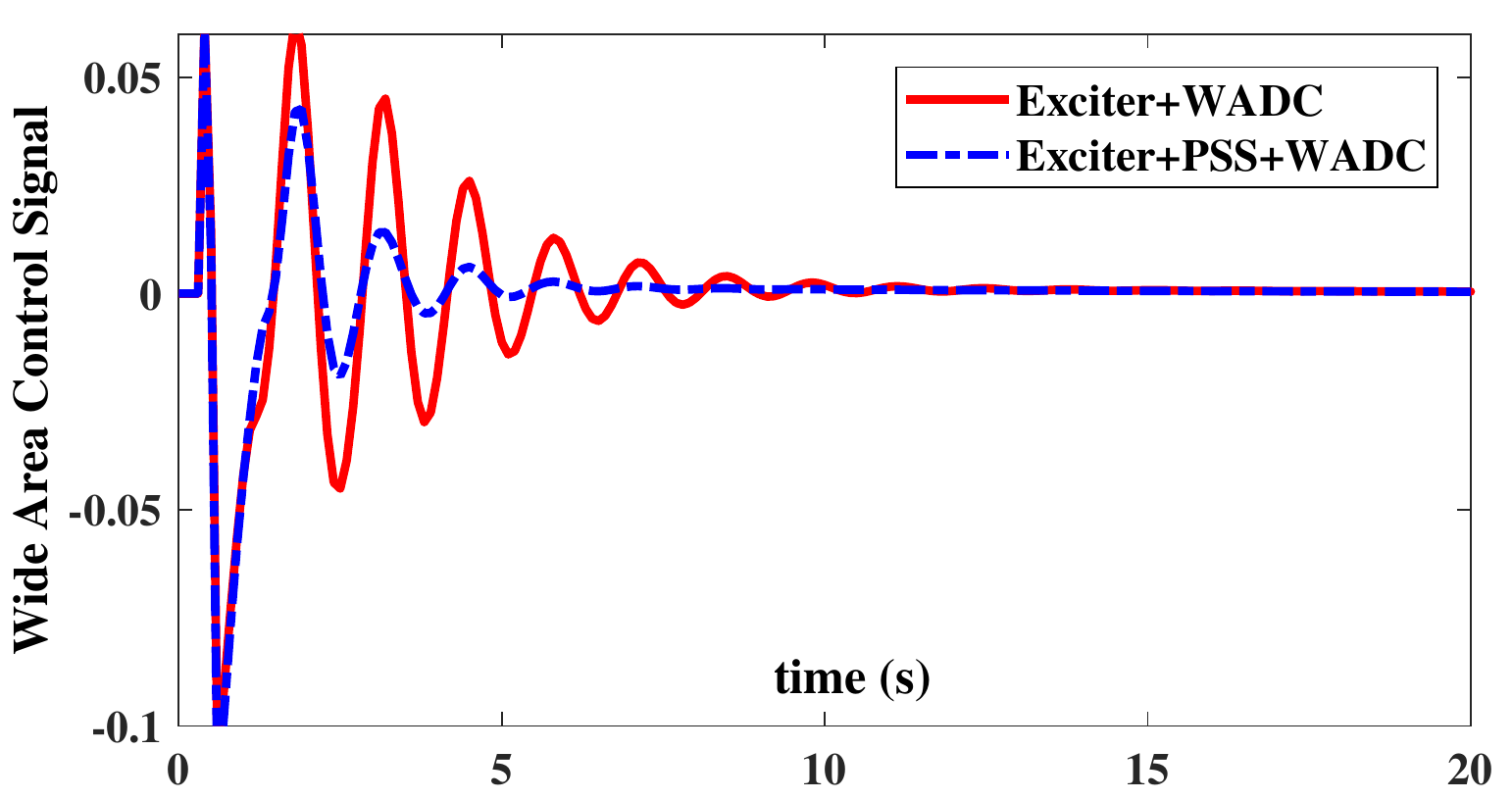}
\caption{Wide area control signal to generator-3.}
\label{fig10aa}
\end{figure}

Table. \ref{table1ab} shows the controller parameters. Fig. \ref{fig7aa} and Fig. \ref{fig8aa} shows the relative speeds of generators-1 and generator-3 with respect to generator-2 (swing) respectively. Fig. \ref{fig9aa} shows the active power flow deviation through tie-lines and Fig. \ref{fig10aa} shows the wide area controller action. Based on the above results it can be seen that WADC is effective in damping inter-area oscillations using the optimal wide area control loop. It can be seen that the proposed approach provides better damping of oscillations.

\begin{table}[!h]
\renewcommand{\arraystretch}{1.3}
\centering
\caption{CONTROLLER PARAMETERS}
\label{table1ab}
\begin{tabular}{*9c} 
\toprule
\shortstack{Controller \\Parameter} & \shortstack{Two-Area \\ System} & \shortstack{39-bus$^\ast$ \\ (Scenario-1)} & \shortstack{39-bus$^\dagger$ \\ (Scenario-2)}\\
\hline
$K_{WADC}$ & -0.5574 & -0.4091 & -0.4599\\
\hline
$T_w$ & 10 & 10 & 10 \\
\hline
$T_{Lead}$ & 0.3253 & 0.4225 & 0.3561\\
\hline
$T_{Lag}$ & 0.1832 & 0.1671 & 0.1960\\
\hline
$V_{min}$ & -0.15 & -0.15 & -0.15\\
\hline
$V_{max}$ & 0.15 & 0.15 & 0.15\\
\hline
\multicolumn{4}{c}{$\ast$=For control loop between Line(Bus15-Bus14) to Gen-6}\\
\multicolumn{4}{c}{$\dagger$=For control loop between Line(Bus39-Bus1) to Gen-10}\\
\hline
\end{tabular}
\end{table}

\subsection{Implementation test results using 39-bus system}
To further validate the proposed algorithm on a larger system and to test the algorithm for various fault scenarios IEEE 39-bus system is used. The effect of selecting a wrong control loop is also analyzed here. The control loops with relatively high residue value compared to other loops are strong loops, whereas control loops with low residue value when compared to other control loops is the weak control loop.
Here WADC (Strong) represents a strong wide-area control loop and WADC (Weak) represents weak wide-area control loop.

\subsubsection{Scenario:1 Fault on Bus-14}
\label{sec22}
\begin{figure}[!h]
\centering
\includegraphics[clip=true, trim=2cm 3.3cm 2.8cm 1.6cm, height=2.5in, width=0.475\textwidth]{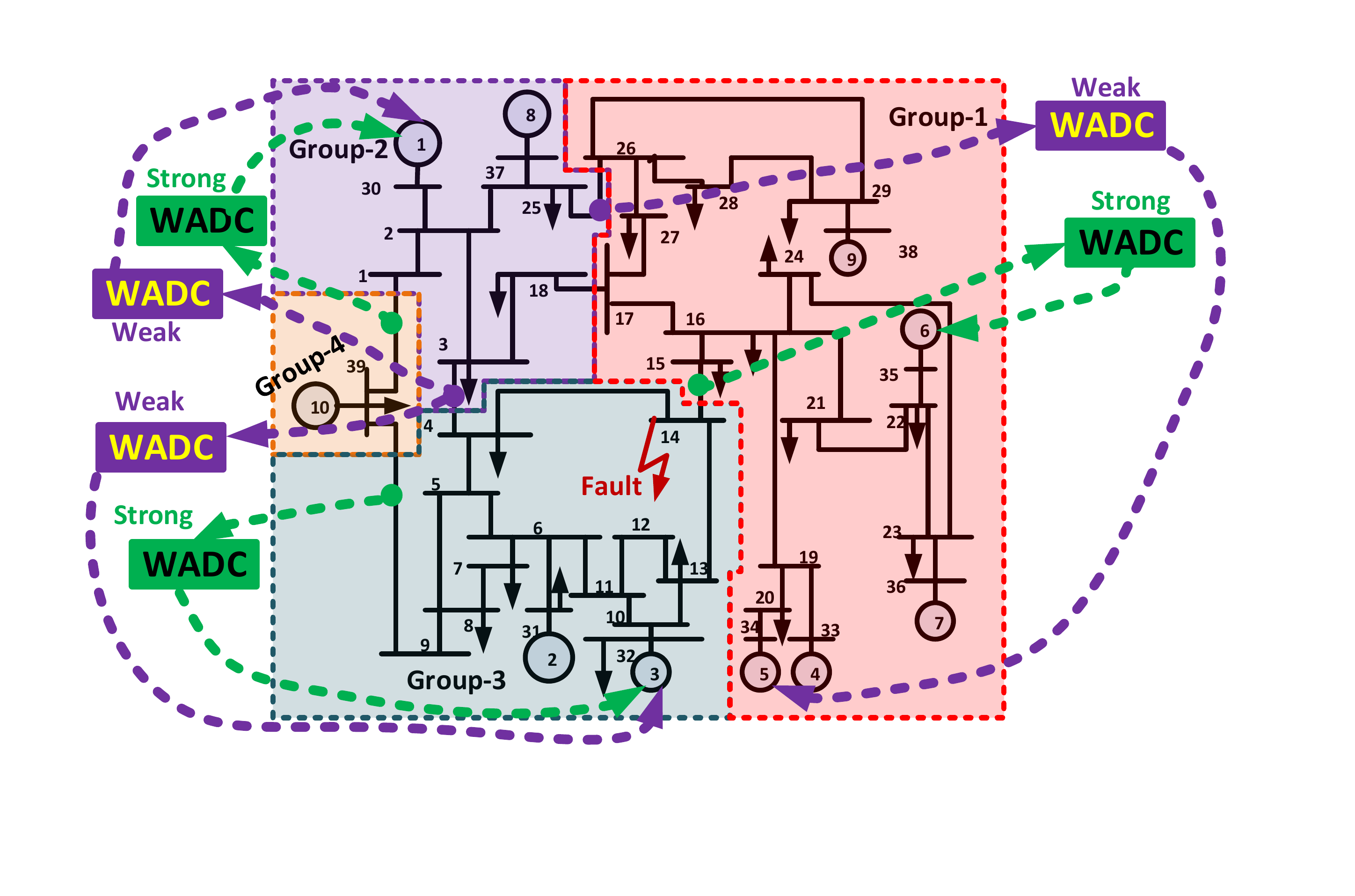}
\caption{Control Loops for IEEE-39 Bus (Bus-14 fault).}
\label{fig14f}
\end{figure}

A three-phase fault is created on Bus-14 for a duration of 0.1 sec at 2 sec. Table. \ref{table_2a} shows the three control loops for each area with high and low residues. Using this information the effect of strong and weaker control loops are analyzed. Fig. \ref{fig14f} shows the strong and weak control loops for a fault on bus-14.

\begin{table}[!t]
\renewcommand{\arraystretch}{1.3}
\centering
\caption{IEEE 39-BUS CONTROL LOOP \\ (FAULT ON BUS-14)}
\label{table_2a}
\begin{tabular}{*9c}
\hline
\toprule
 & Observable & Controllable & Residue \\
\hline
\multirow{3}{*}{High} & Line (Bus15-Bus14) & Gen-6 & 5.5871\\
 & Line (Bus9-Bus39) & Gen-3 & 4.6915\\
 & Line (Bus1-Bus39) & Gen-1 & 2.4572\\
\hline
\multirow{3}{*}{Low} & Line (Bus26-Bus25) & Gen-5 & 0.2753\\
 & Line (Bus4-Bus3) & Gen-3 & 0.1512\\
 & Line (Bus3-Bus4) & Gen-1 & 0.1175\\
\hline
\end{tabular}
\end{table}

\begin{figure}[!h]
\centering
\includegraphics[width=0.475\textwidth]{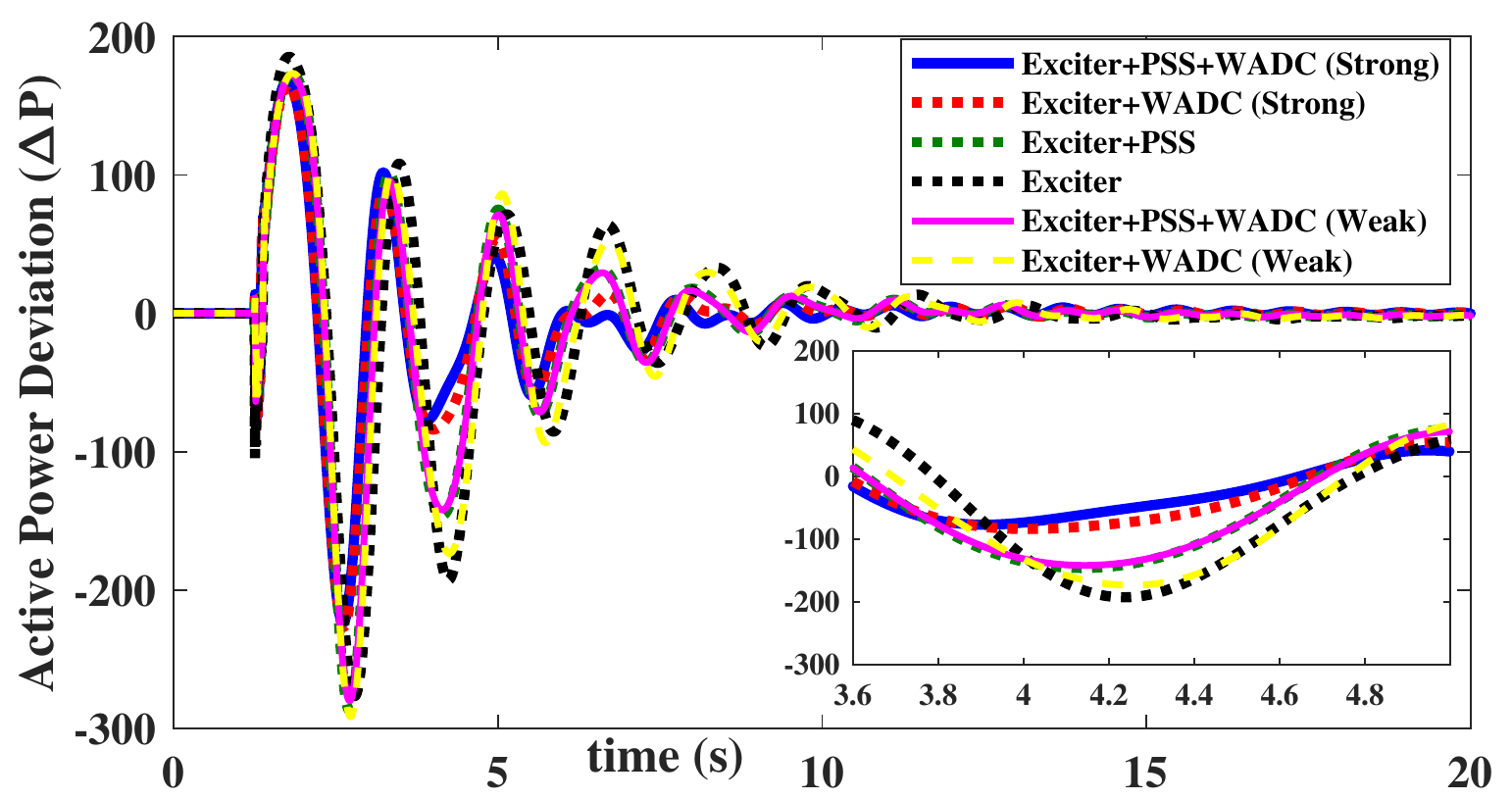}
\caption{Active power deviation (Bus17-Bus18).}
\label{fig7x}
\end{figure}

\begin{figure}[!h]
\centering
\includegraphics[width=0.475\textwidth]{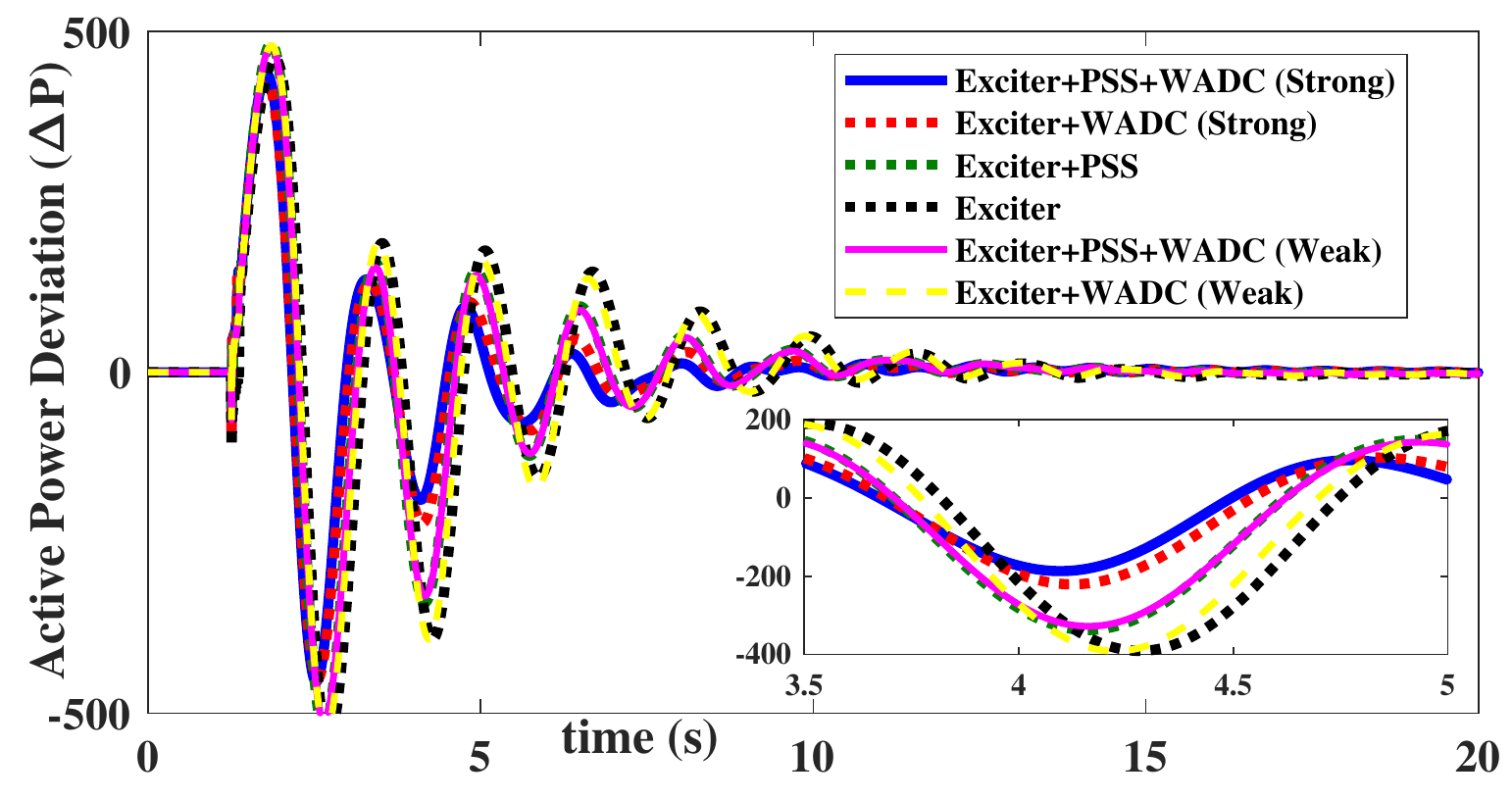}
\caption{Active power deviation (Bus1-Bus39).}
\label{fig8x}
\end{figure}

\begin{figure}[!h]
\centering
\includegraphics[width=0.475\textwidth]{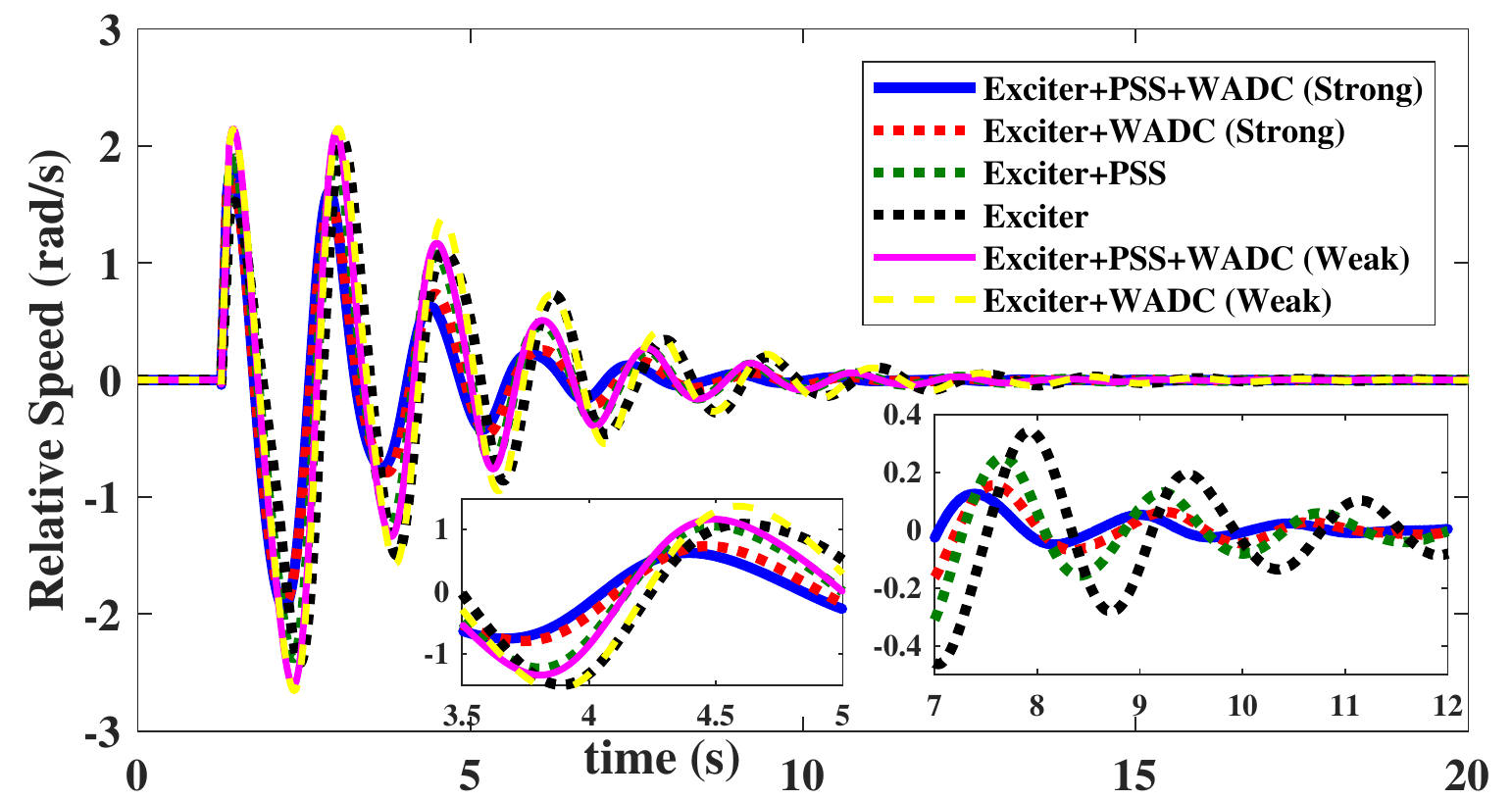}
\caption{Relative speed between generator 2 and generator 4.}
\label{fig9x}
\end{figure}

\begin{figure}[!h]
\centering
\includegraphics[width=0.475\textwidth]{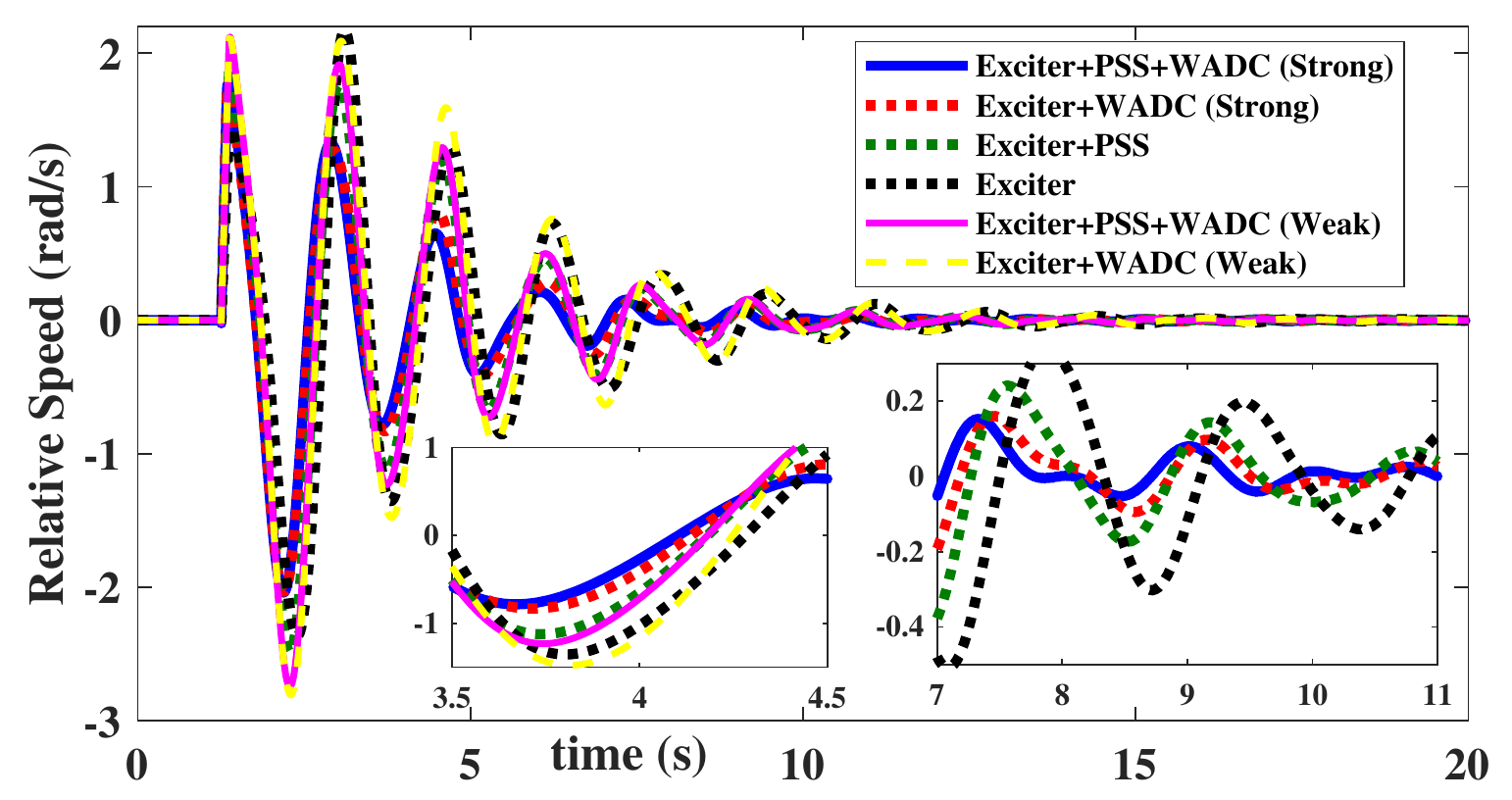}
\caption{Relative speed between generator 2 and generator 6.}
\label{fig10x}
\end{figure}

\begin{figure}[!h]
\centering
\includegraphics[width=0.475\textwidth]{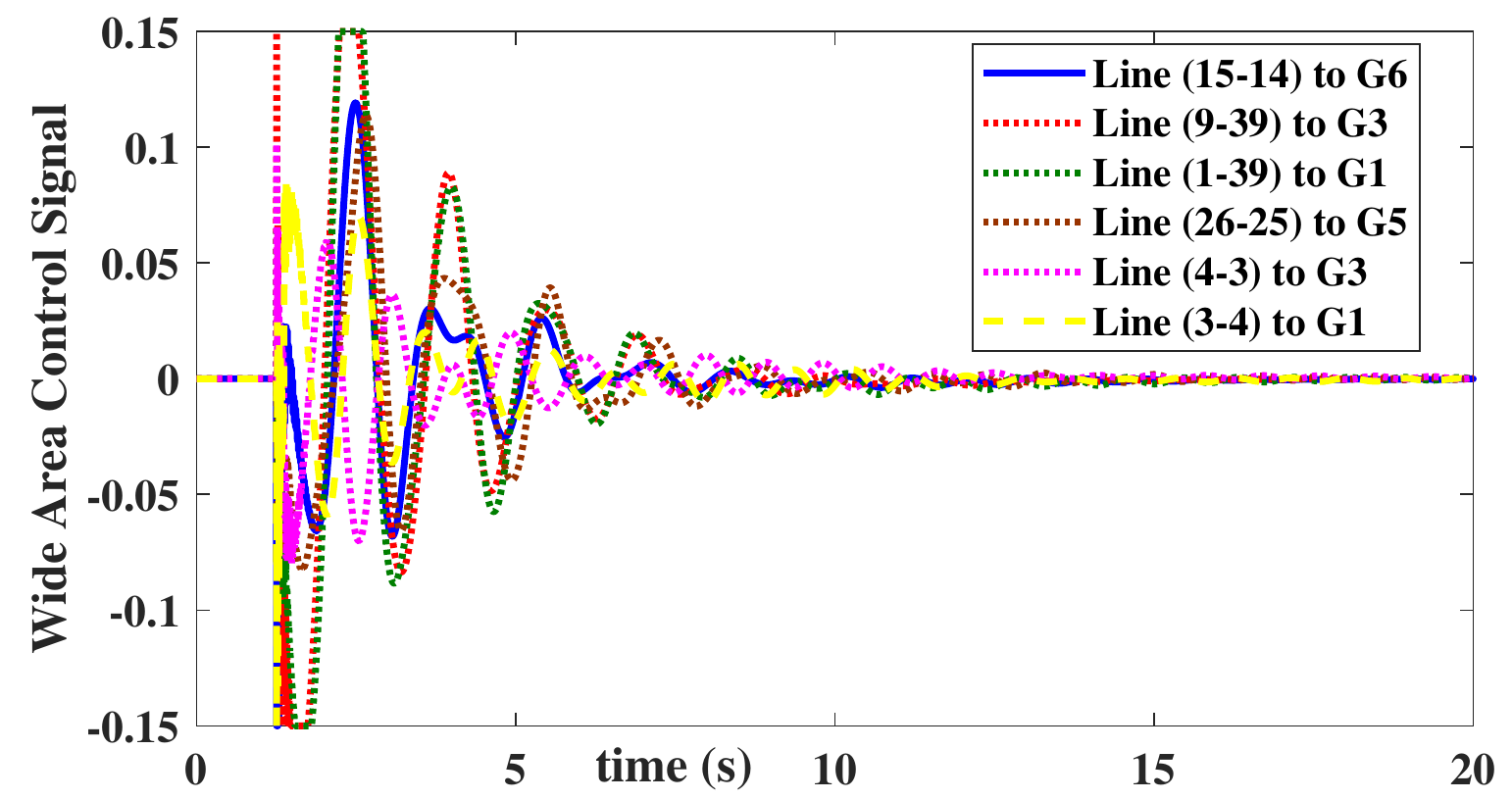}
\caption{Wide area control output with PSS.}
\label{fig11x}
\end{figure}

Fig. \ref{fig7x} and Fig. \ref{fig8x} shows the active power deviation of the lines connected between Bus17-Bus18 and Bus1-Bus39 respectively. Fig. \ref{fig9x} and Fig. \ref{fig10x} shows the relative speed between generator 4 and generator 6 w.r.t swing generator-2 respectively. Fig. \ref{fig11x} shows the wide area control output for the case with PSS. Table. \ref{table_comp14} shows the active power deviation and relative speed at a sample point (trough of oscillation here). From the Table \ref{table_comp14} it can be seen that when compared to Exciter only case, addition of a PSS reduced oscillations by 29.52\%, addition of WADC (Strong) reduce the oscillation by 63.54\%, addition of PSS and WADC (Strong) reduce the oscillation by 75.45\%, with the addition of WADC (Weak) the oscillations reduce by 8.4\%, and with the addition of PSS and WADC (Weak) the oscillations reduce by 30.93\%. The relative speed oscillations with PSS, WADC (Strong), PSS and WADC (Strong), WADC (Weak), PSS and WADC (Weak) are reduced by 24.61\%, 57.66\%, 68.77\%, 2.55\%, and 24.65\% respectively. It can be concluded that with WADC the oscillations are damped much more effectively. It can also be observed that with WADC the oscillations are damped more effectively especially if the optimal control loop is strong. 

\begin{table}[!t]
\renewcommand{\arraystretch}{1.3}
\centering
\caption{CONTROLLER PERFORMANCE COMPARISON \\ (FAULT ON BUS-14)}
\label{table_comp14}
\begin{tabular}{*9c}
\hline
\toprule
\shortstack{Controller \\ \ } & \shortstack{Active Power \\ Deviation (MW) \\ at 4.3s (Fig. \ref{fig7x})} & \shortstack{Relative Speed \\ (rad/s) \\ at 3.9s (Fig. \ref{fig9x})} \\
\hline
Exciter & -187.54 & -1.4952 \\
Exciter+PSS & -132.1754 & -1.1272 \\
Exciter+WADC (Strong) & -46.044 & -0.6330 \\
Excite+PSS+WADC (Strong) & -68.37 & -0.4670 \\
Exciter+WADC (Weak) & -171.7852 & -1.4570 \\
Exciter+PSS+WADC (Weak) & -129.5395 & -1.1267 \\
\hline
\end{tabular}
\end{table}

\subsubsection{Scenario:2 Fault on Bus-25}
 In this scenario a three phase fault is created on Bus-25 for a duration of 0.1 sec at 2 sec. Table \ref{table_3b} shows the three control loops for each area with high and low residues. Using this information the effect of strong and weaker control loops are analyzed.
\begin{table}[!t]
\renewcommand{\arraystretch}{1.3}
\centering
\caption{IEEE 39-BUS CONTROL LOOP \\ (FAULT ON BUS-25)}
\label{table_3b}
\begin{tabular}{*9c}
\hline
\toprule
 & Observable & Controllable & Residue \\
\hline
\multirow{3}{*}{High} & Line (Bus39-Bus1) & Gen-10 & 7.8621\\
 & Line (Bus39-Bus9) & Gen-10 & 6.9582\\
 & Line (Bus1-Bus39) & Gen-8 & 4.1673\\
\hline
\multirow{3}{*}{Low} & Line (Bus15-Bus14) & Gen-5 & 0.5836\\
 & Line (Bus26-Bus25) & Gen-9 & 0.3949\\
 & Line (Bus4-Bus3) & Gen-2 & 0.3936\\
 
\hline
\end{tabular}
\end{table}

\begin{figure}[!h]
\centering
\includegraphics[width=0.475\textwidth]{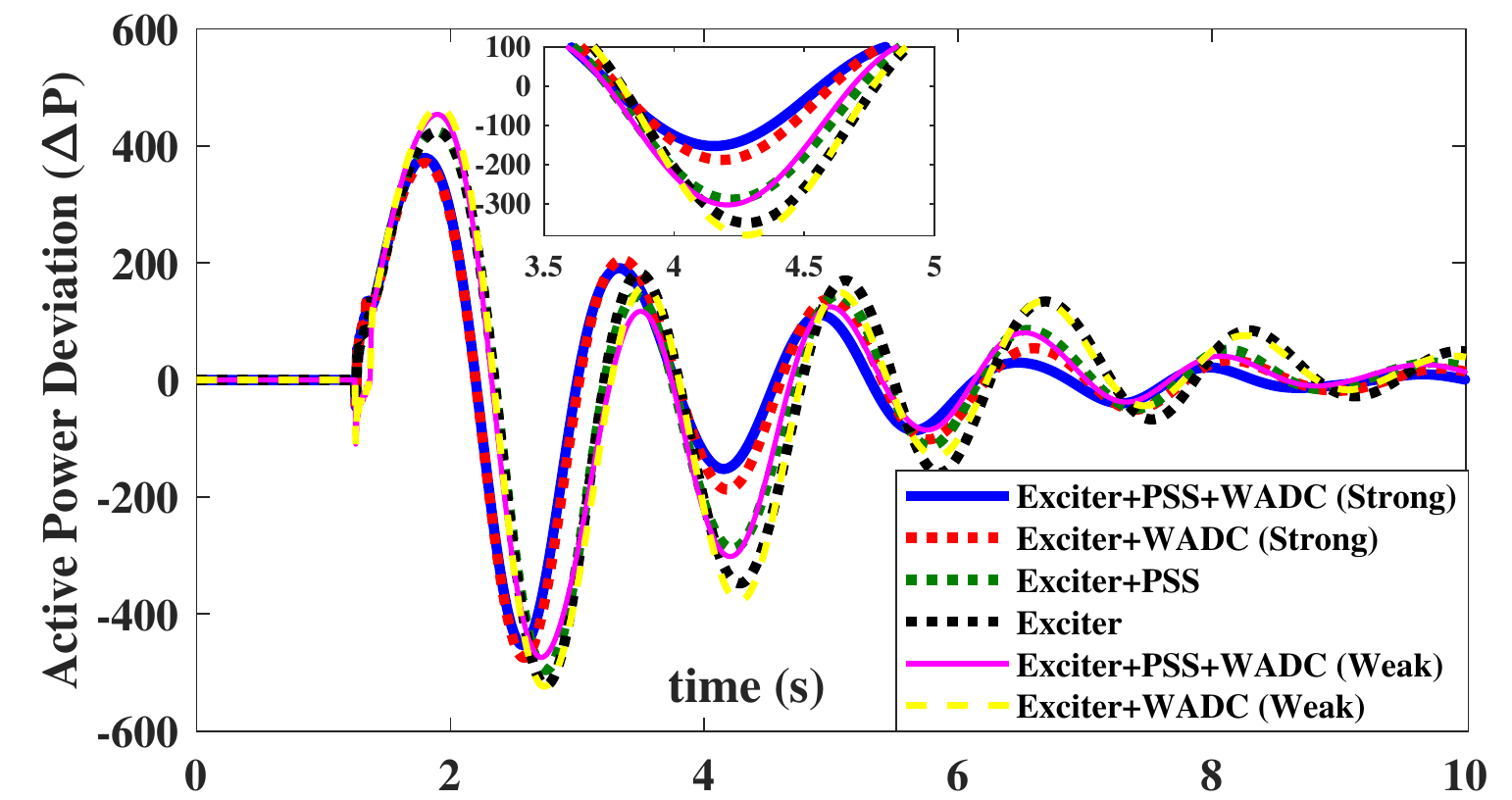}
\caption{Active power deviation (Bus1-Bus39).}
\label{fig1x}
\end{figure}

\begin{figure}[!h]
\centering
\includegraphics[width=0.475\textwidth]{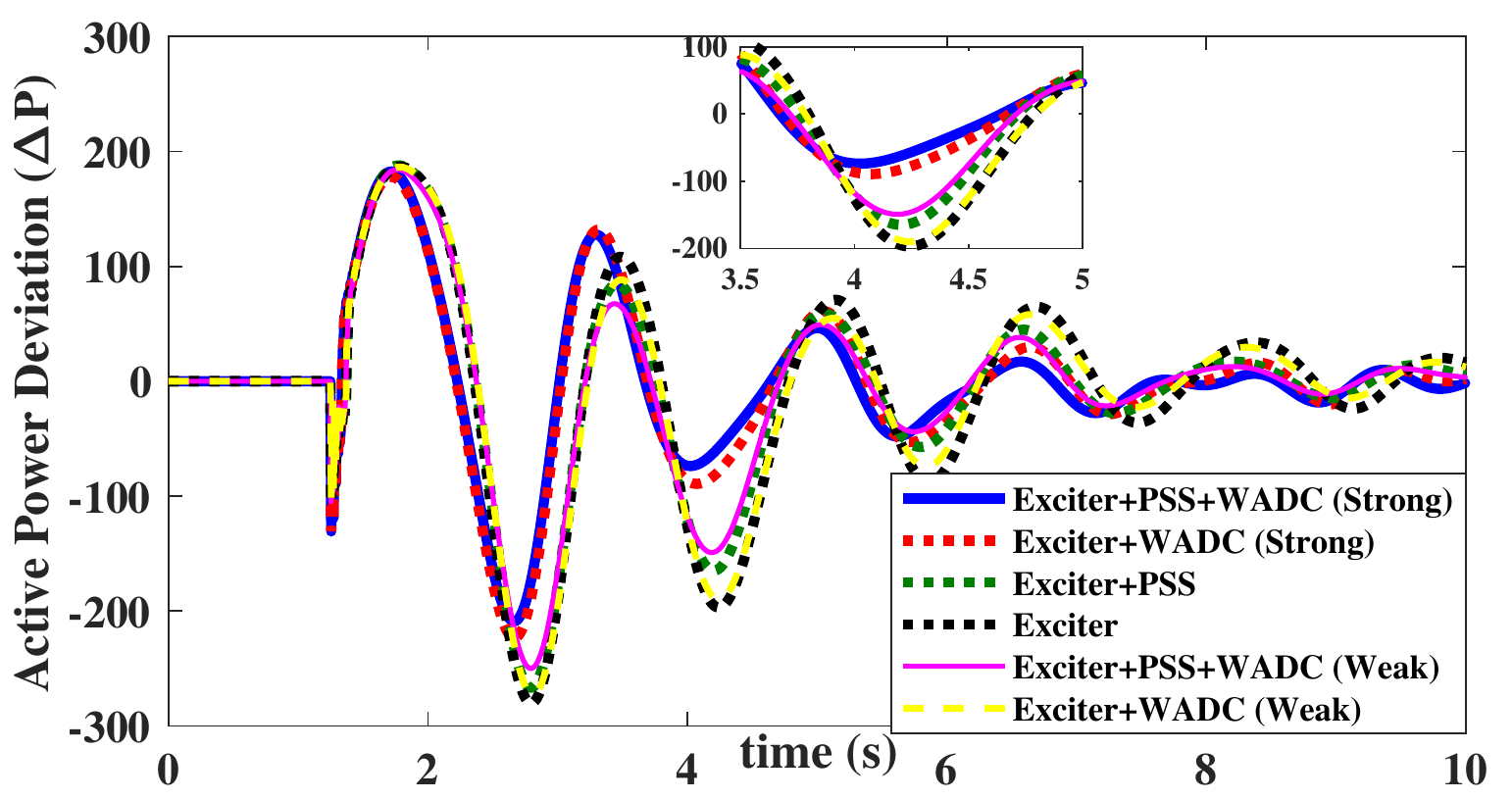}
\caption{Active power deviation (Bus17-Bus18).}
\label{fig2x}
\end{figure}

\begin{figure}[!h]
\centering
\includegraphics[width=0.475\textwidth]{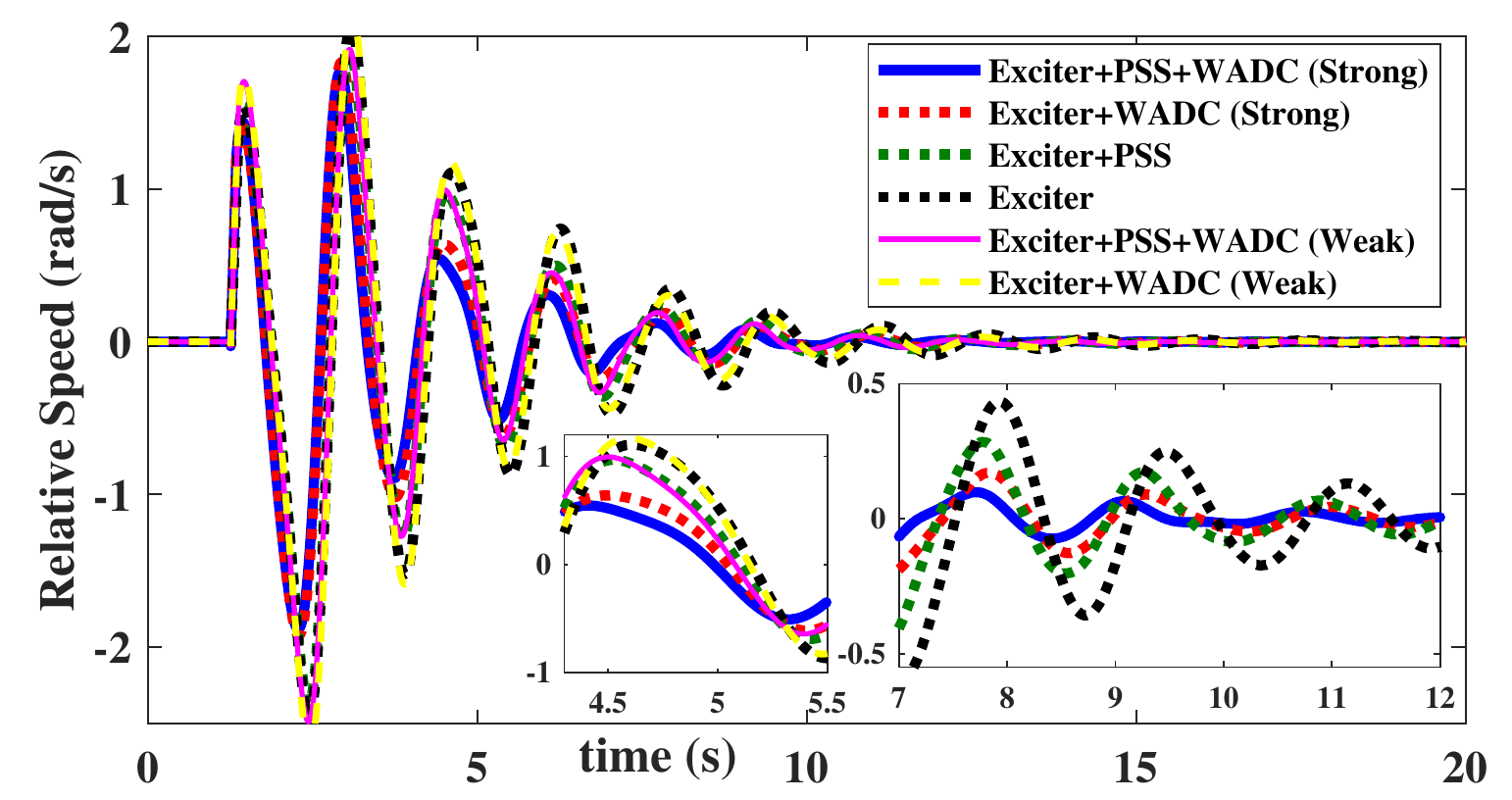}
\caption{Relative speed between generator 2 and generator 4.}
\label{fig3x}
\end{figure}

\begin{figure}[!h]
\centering
\includegraphics[width=0.475\textwidth]{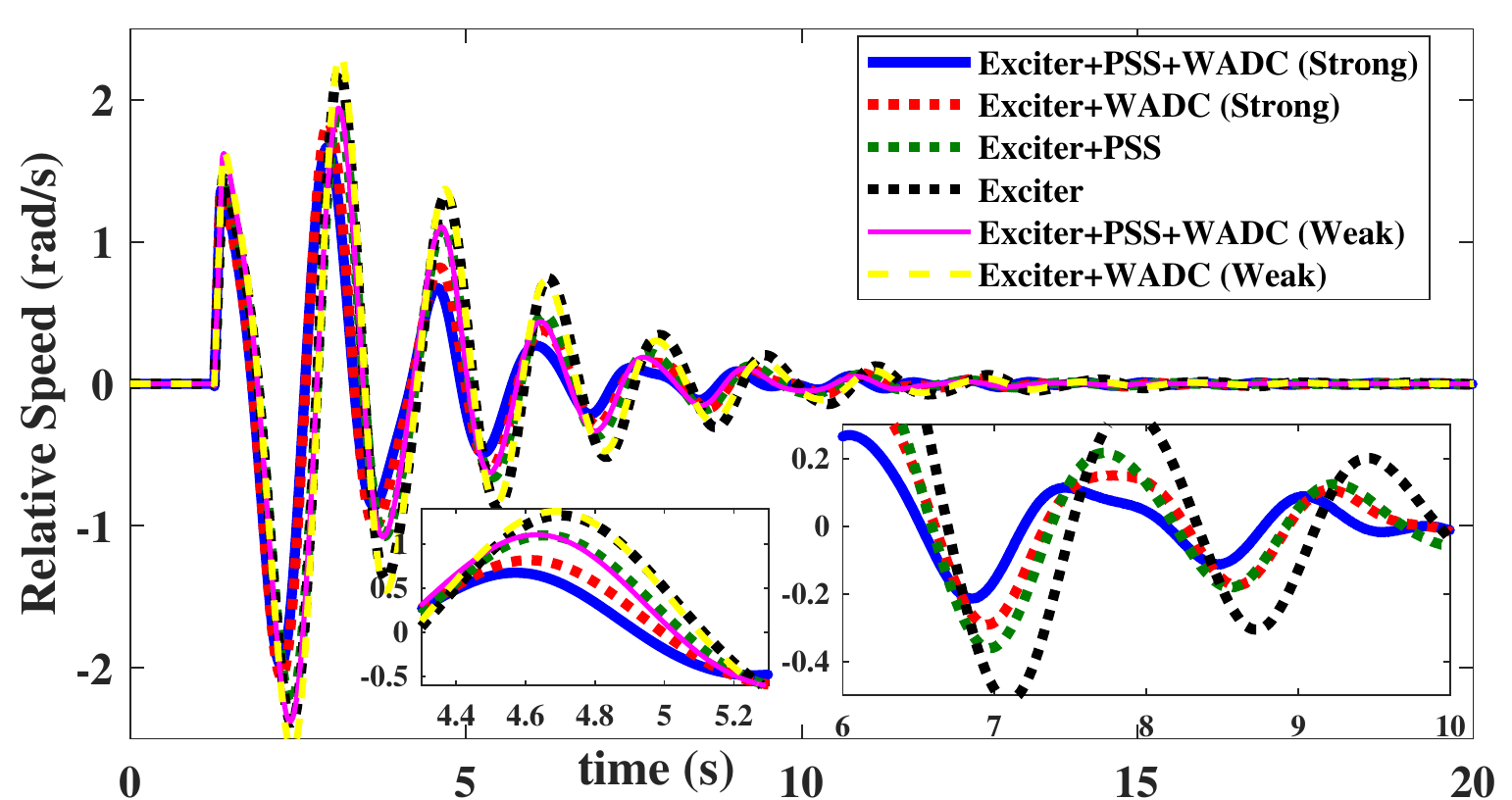}
\caption{Relative speed between generator 2 and generator 6.}
\label{fig4x}
\end{figure}

\begin{figure}[!h]
\centering
\includegraphics[width=0.475\textwidth]{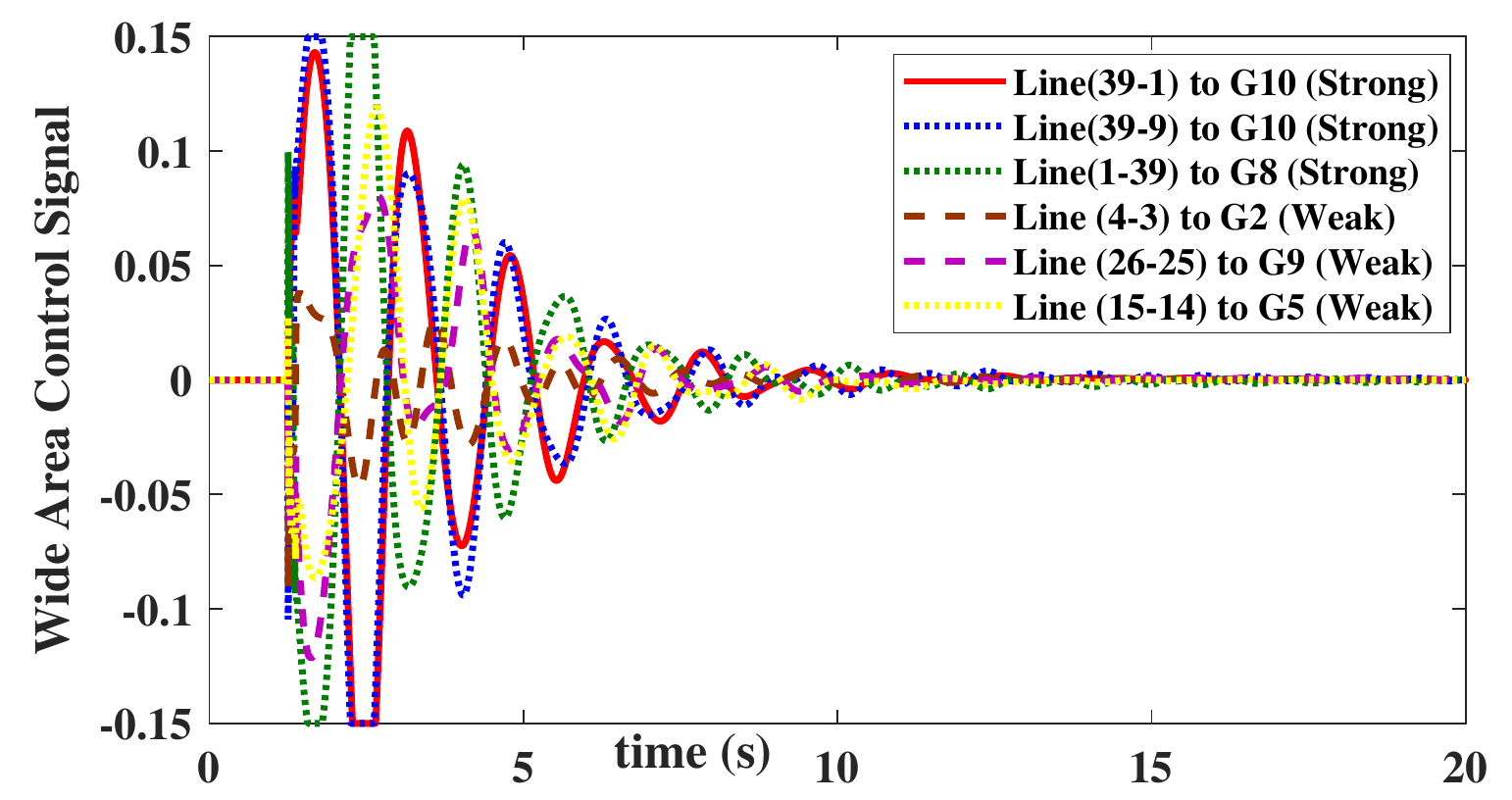}
\caption{Wide area control output with PSS.}
\label{fig5x}
\end{figure}

Fig. \ref{fig1x} and Fig. \ref{fig2x} shows the active power deviation of the lines connected between Bus1-Bus39 and Bus17-Bus18 respectively. Fig. \ref{fig3x} and Fig. \ref{fig4x} shows the relative speed between generator 4 and generator 6 w.r.t swing generator-2 respectively. Fig. \ref{fig5x} shows the wide area control output for the case with PSS. Table \ref{table_comp25} shows the active power deviation and relative speed at a sample point (trough of oscillation here). From the Table \ref{table_comp25} it can be seen that when compared to Exciter only case, addition of a PSS reduced oscillations by 19.68\%, addition of WADC (Strong) reduce the oscillation by 63.26\%, addition of PSS and WADC (Strong) reduce the oscillation by 74.03\%, with the addition of WADC (Weak) the oscillations reduce by 3.4\%, and with the addition of PSS and WADC (Weak) the oscillations reduce by 28.62\%. The relative speed oscillations with PSS, WADC (Strong), PSS and WADC (Strong), WADC (Weak), PSS and WADC (Weak) are reduced by 17.10\%, 51.97\%, 62.5\%, 2.01\%, and 18.94\% respectively. It can be concluded that with WADC the oscillations are damped much more effectively. As before, it can also be seen that with WADC the oscillations are damped more effectively if the optimal control loop is strong.

\begin{table}[!t]
\renewcommand{\arraystretch}{1.3}
\centering
\caption{CONTROLLER PERFORMANCE COMPARISON \\ (FAULT ON BUS-25)}
\label{table_comp25}
\begin{tabular}{*9c}
\hline
\toprule
\shortstack{Controller \\ \ } & \shortstack{Active Power \\ Deviation (MW) \\ at 4.3s (Fig. \ref{fig1x})} & \shortstack{Relative Speed \\ (rad/s) \\ at 3.9s (Fig. \ref{fig3x})} \\
\hline
Exciter & -192.39 & -1.52 \\
Exciter+PSS & -154.52 & -1.26 \\
Exciter+WADC (Strong) & -70.67 & -0.73 \\
Excite+PSS+WADC (Strong) & -49.95 & -0.57 \\
Exciter+WADC (Weak) & -185.78 & -1.4894 \\
Exciter+PSS+WADC (Weak) & -137.3218 & -1.2321 \\
\hline
\end{tabular}
\end{table}

Further to compare the effectiveness of the proposed algorithm, the norms of different oscillation cases are compared with exciter only case. The relative error metric is given by \eqref{eqn44}. 
\begin{equation}
\label{eqn44}
relative\ error = \frac{\left\|y_{exc}-y_{act}\right\|_2}{\left\|y_{exc}\right\|_2}
\end{equation}
where $y_{exc}$ is the data with exciter only, $y_{act}$ is the data with other cases as shown in Table \ref{table_2no}, and $\left\|\ \right\|$ is the 2-norm of a vector. Using the relative error metric it can be observed that the larger the error between with exciter only case and other cases with supplementary controls like PSS and WADC, more effective is the damping.

\begin{table}[!t]
\renewcommand{\arraystretch}{1.3}
\centering
\caption{RELATIVE ERROR COMPARISON}
\label{table_2no}
\begin{tabular}{*9c}
\hline
\toprule
Variable & Case-1 & Case-2 & Case-3 & Case-4 & Case-5\\
\hline
Fig. \ref{fig9x} & 0.7146 & 0.6732 & 0.4107 & 0.4230 & 0.3137\\
Fig. \ref{fig10x} & 0.7587 & 0.6667 & 0.4232 & 0.4403 & 0.3\\
Fig. \ref{fig7x} & 0.6831 & 0.6014 & 0.3773 & 0.3704 & 0.2327 \\
Fig. \ref{fig8x} & 0.6552 & 0.5687 & 0.3388 & 0.3368 & 0.1733 \\
\hline
Fig. \ref{fig3x} & 0.6228 & 0.5519 & 0.2061 & 0.2543 & 0.1013 \\
Fig. \ref{fig4x} & 0.6254 & 0.5391 & 0.1998 & 0.2545 & 0.1021 \\
Fig. \ref{fig1x} & 0.6093 & 0.54061 & 0.1938 & 0.2728 & 0.1003 \\
Fig. \ref{fig2x} & 0.6004 & 0.5367 & 0.1906 & 0.2125 & 0.1320 \\
\hline
\multicolumn{6}{c}{Case1: Exciter+PSS+WADC (Strong), Case2: Exciter+WADC (Strong)}\\
\multicolumn{6}{c}{Case3: Exciter+PSS, Case4: Exciter+WADC (Weak)}\\
\multicolumn{6}{c}{Case5: Exciter+PSS+WADC (Weak)}\\
\bottomrule
\end{tabular}
\end{table}

\subsection{Other Scenarios}
For extensive validation of the proposed algorithm, various other scenarios with a disturbance in different locations are also analyzed. The percentage reduction in oscillations with supplementary control (Exciter+PSS+WADC) when compared to Exciter only case is summarized in Table \ref{table_ex}.

\begin{table}[!t]
\renewcommand{\arraystretch}{1.3}
\centering
\caption{PERCENTAGE REDUCTION IN OSCILLATIONS}
\label{table_ex}
\begin{tabular}{*9c}
\hline
\toprule
\multirow{2}{*}{Disturbance Type} & \multicolumn{2}{c}{\% Reduction in Oscillations} \\
\cline{2-3}
 & Active Power Deviation & Relative Speed \\
\hline
Fault on Bus-4 & 71.17\% & 60.49\%\\
\hline
Fault on Bus-18 & 75.82\% & 64.15\%\\
\hline
Load Drop on Bus-3 & 79.67\% & 72.86\%\\
\hline
Load Drop on Bus-16 & 77.95\% & 71.37\%\\
\hline
\bottomrule
\end{tabular}
\end{table}

\subsection{Effect of Time-Delay}
\begin{figure}[!h]
\centering
\includegraphics[width=0.475\textwidth]{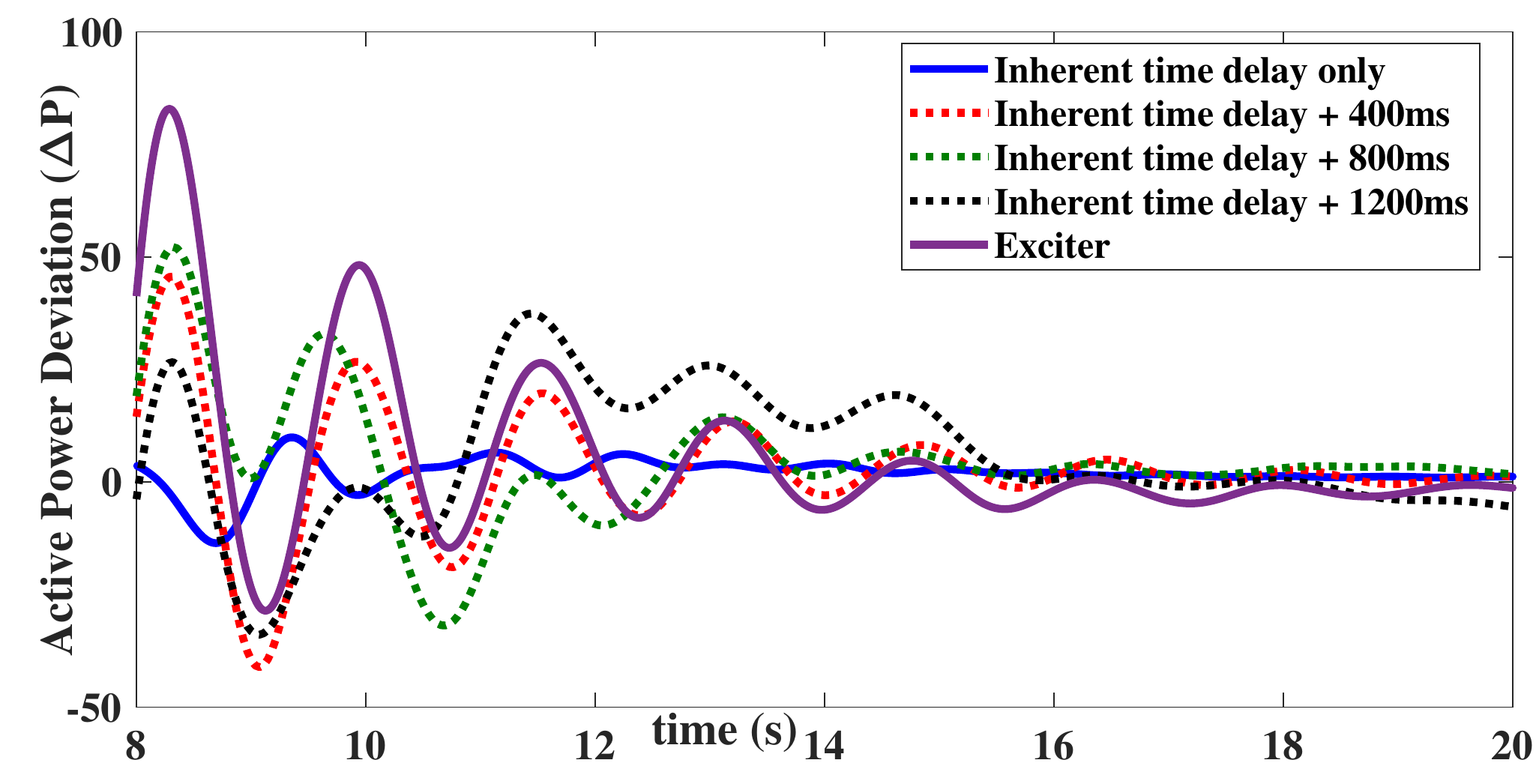}
\caption{Active power flow deviation (Bus-9 to Bus-39)}
\label{fig6x}
\end{figure}

To study the effect of time delay on the control design process, an intentional time-delay is added to the control process in addition to inherent time-delay. For this, a test case as discussed in Section \ref{sec22} is re-simulated with various time delays. Fig. \ref{fig6x} shows the Line (Bus-9 to Bus-39) active power flow deviation with strong wide-area control loop and exciter, PSS and WADC activated and with Exciter only. It can be observed that with time delay of $400ms$ (limit imposed by standards) the oscillations are similar or less than that of exciter case. It can also be seen that with an unrealistic delay of $1200ms$, the system is stills stable with less first swings when compared to exciter case. 
\vspace{-8mm}

\subsection{Limitations/Assumptions and Future Directions}
For the design of the proposed algorithm following assumptions are made.

\begin{description}
  \item[$\bullet$] \textbf{Dynamic coherency grouping}:
In this paper areas are pre-defined based on coherency grouping, however the changing coherency groups with changes in operating conditions can be considered in the control action for more effective performance.
\item[$\bullet$] \textbf{Data Synchronization}:
In this paper it is assumed that data from all the areas are processed at the same time, however, there may be some delay for this time-stamping of data and asynchronous ADMM \cite{AADMM} can be used to address the delay problems.
\item[$\bullet$] \textbf{Use of NS-3 network simulator}:
For emulating cyber security scenarios on the communication side NS-3 network simulator \cite{ns3} can be used.
\end{description}
 
\section{Conclusions}
The proposed method for selection of optimal control loop for wide area control using ADMM based distributed algorithms overcome the drawbacks of earlier methods reported in the literature. In this approach, the interconnected power system is divided into areas, and then using measurements a multi-input-multi-output (MIMO) black-box transfer function model is estimated locally for each area based on the Lagrange multipliers method. The local area processors communicates with the global processor to estimate a global transfer function model of the power system. The information of residue corresponding to inter-area mode obtained from global transfer function is used for selecting an optimal wide area control loop and to design WADC. The efficacy of the proposed distributed approach is validated using two-area and IEEE 39 bus power system models on RTDS/RSCAD® and MATLAB® co-simulation platform. From the simulation results it is found that the proposed distributive method effectively damp the  inter-area oscillations (up to 80\%) effectively.


%

\ifCLASSOPTIONcaptionsoff
  \newpage
\fi

\bibliographystyle{IEEEtran}
\bibliography{Ref.bbl}

\vspace{-13 mm}
\begin{IEEEbiography}[{\includegraphics[width=1in,height=1.25in,clip,keepaspectratio]{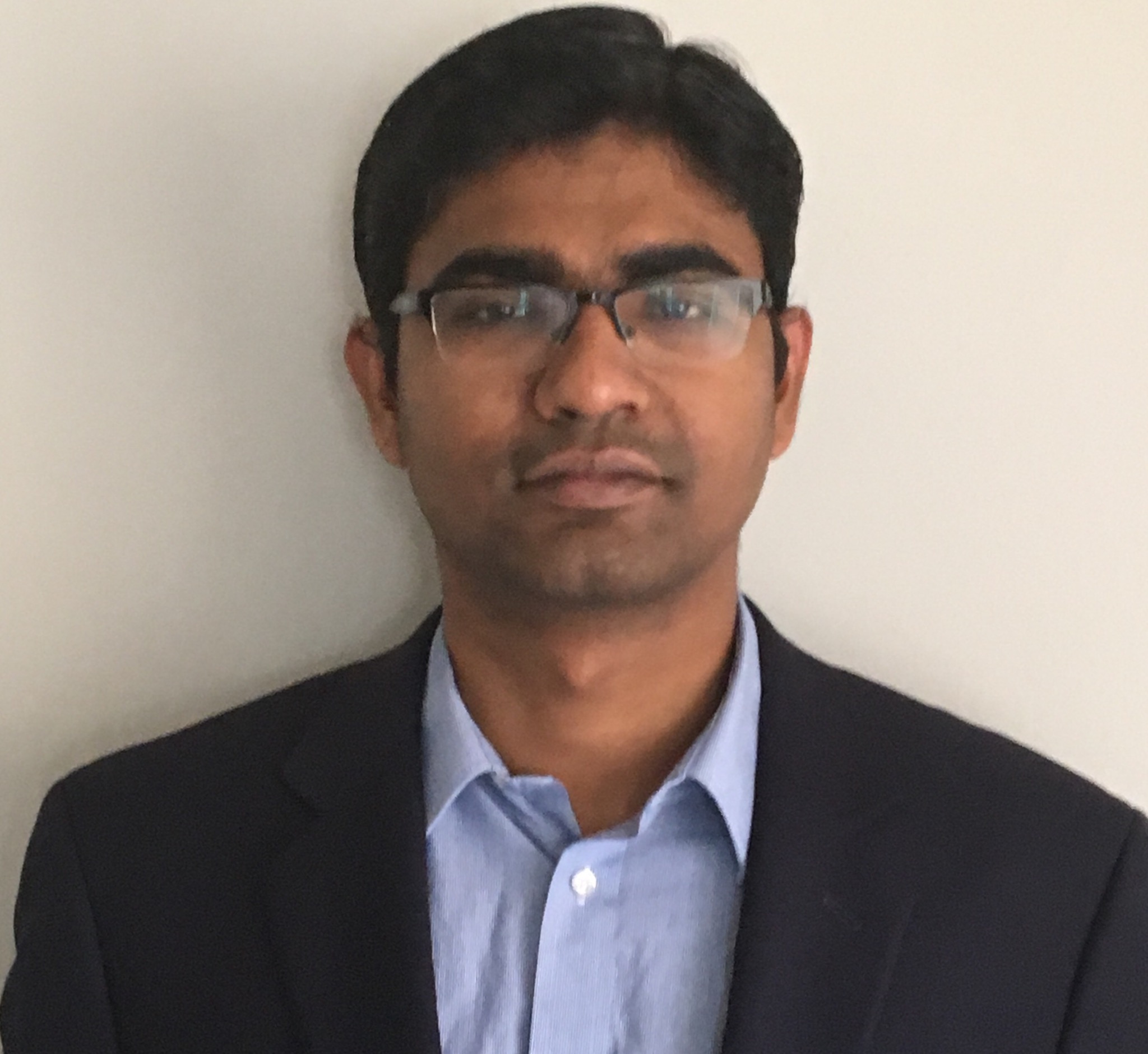}}]{A. Thakallapelli}(S'14, M'19) received his B.Tech degree in Electrical Engineering from Acharya Nagarjuna University in 2010, M.Tech degree in Electrical Engineering from the Veermata Jijabai Technological Institute in 2012, and Ph.D degree in Electrical Engineering from the Department of Electrical and Computer Engineering, University of North Carolina at Charlotte in 2018. His research interests include wide-area control, reduced order modeling, power system stability and renewable energy.
\end{IEEEbiography}
\vspace{-13 mm}
\begin{IEEEbiography}
[{\includegraphics[width=1in,height=1.25in,clip,keepaspectratio]{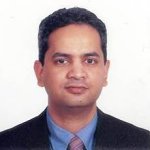}}]{S. Kamalasadan}(S'01, M'05, SM'17) received his B Tech. degree in Electrical and Electronics from the University of Calicut, Kerala, India in 1991, M.Eng in Electrical Power Systems Management, from the Asian Institute of Technology, Bangkok, Thailand in 1999, and Ph.D. in Electrical Engineering from the University of Toledo, Ohio, USA in 2004. He is currently working as a Professor in the department of electrical and computer engineering at the University of North Carolina at Charlotte. He has won several awards including the NSF CAREER award and IEEE best paper award. His research interests include Intelligent and Autonomous Control, Power Systems dynamics, Stability and Control, Smart Grid, Micro-Grid and Real-time Optimization and Control of Power System. 
\end{IEEEbiography}

\end{document}